\documentclass[11pt]{article}
\usepackage[makeroom]{cancel}
\usepackage{cite}
\usepackage{graphicx}
\usepackage{multirow}
\usepackage[utf8]{inputenc}
\usepackage[english]{babel}
\usepackage{graphicx}
\usepackage{amsmath}
\usepackage{multicol,lipsum,graphicx,float}
\usepackage{color}
\usepackage{caption}
\usepackage{cite}
\usepackage{blindtext}
\usepackage{hyperref}
\usepackage{epstopdf}
\usepackage[bottom]{footmisc}
\makeatletter
\newcommand*{\rom}[1]{\expandafter\@slowromancap\romannumeral #1@}
\makeatother
\begin{document}
\begin{center}
{\Large\bf  Dark Matter as a remnant of  SQCD Inflation}
%\\[.2mm]
%\vskip .2cm
\\
\vskip .5cm
{ Subhaditya Bhattacharya$^{a,}$\footnote{subhaditya@iitg.ac.in}
Abhijit Kumar Saha$^{a,}$\footnote{abhijit.saha@iitg.ac.in},
Arunansu Sil$^{a,}$\footnote{asil@iitg.ac.in}, Jose Wudka$^{b,}$\footnote{jose.wudka@ucr.edu}}\\[3mm]
{\it{
$^a$ Indian Institute of Technology Guwahati, 781039 Assam, India,}\\
$^b$ University of California, Riverside, California 92521, USA
}
\end{center}

\vskip .5cm

\begin{abstract}
We propose a strongly coupled supersymmetric gauge theory that can accommodate 
both the inflation (in the form of generalized hybrid inflation) and dark matter (DM). In this 
set-up, we identify the DM as the Goldstones associated with the breaking 
of a global symmetry ($SU(4)\times SU(4) \to SU(4)$) after inflation ends. 
Due to the non-abelian nature of this symmetry, the scenario provides with
multiple DMs. We then construct a low energy theory which generates 
a Higgs portal like coupling of the DMs with Standard Model (SM), thus allowing them 
to thermally freeze out. While the scales involved in the inflation either have a dynamical origin or related 
to UV interpretation in terms of a heavy quark field in the supersymmetric QCD 
(SQCD) sector, the DM masses however are generated from explicit breaking 
of the chiral symmetry of the SQCD sector. We discuss DM phenomenology 
for both degenerate and non-degenerate cases, poised with DM-DM interactions and 
find allowed region of parameter space in terms of relic density and direct search constraints.      

\noindent \hspace{4cm}
\end{abstract}
%%%%%%%%%%%%%%%%%%%%%%%%%%%%%%%%%%%%
\section{Introduction}
\label{sec:intro}
%%%%%%%%%%%%%%%%%%%%%%%%%%%%%%%%%%%%

Amidst the great success of the Standard Model (SM) of particle physics, there are 
several intriguing issues which indicate that the SM should be extended or 
supplemented by some other sector (including new fields and/or gauge  symmetry) 
in order to provide a more complete description of nature. In particular, SM fails to 
accommodate a large share of energy density of the universe (25$ \%$), called 
dark matter (DM). On the other hand, the idea of primordial inflation serves as an
elegant construction which can actually resolve some of the intricate problems 
($e.g$ horizon and flatness problems) of otherwise quite successful Big Bang 
cosmology. This inflationary hypothesis is further strengthened by its prediction 
on the primordial perturbation that leads to striking agreements with the 
observation of the cosmic microwave background radiation (CMBR) spectrum.  
However, SM alone can not be responsible for such a primordial inflation. In this 
work, we consider the existence of a different sector other than the SM, 
which can address primordial inflation and at the same time provides  a suitable 
DM candidate. In \cite{Boucenna:2014uma}, such an attempt to connect between primordial inflation and DM succesfully has been 
made.

A successful model for inflation demands the existence a very flat potential for 
slow-roll conditions to be satisfied during the inflationary epoch. Inclusion of supersymmetry 
protects the flatness of the scalar potential by non-renormalization theorem and 
is therefore a natural possibility. An inflation model embedded 
in a supersymmetric framework (for $e.g.$, supersymmetric hybrid inflation models) 
usually contains one or more mass scales that are quite large compared to the electroweak scale, 
although smaller than the Planck scale as indicated by PLANCK\cite{Ade:2015xua} and WMAP data\cite{Spergel:2006hy}. 
In a remarkable attempt to address the issue\cite{Brax:2009yd}, it was shown that the 
inclusion of a hidden sector in the form of supersymmetric 
QCD (SQCD) can dynamically generate the scale of inflation, or relate it 
to the heavy quark mass of the electric  theory ({\it i.e.} the theory at scales above the strong coupling scale). 
Inflation based on strongly coupled supersymmetric gauge theories has been studied
in \cite{Domcke:2017rzu,Harigaya:2017jny,Domcke:2017xvu,Harigaya:2014wta,Harigaya:2014sua,Harigaya:2012pg,Schmitz:2016kyr}.
The properties of inflation in the SQCD framework are determined by the number of colors ($N_C$) and number of flavors ($N_f$) in the model. 
If one initially chooses  $N_f = N_C + 1$ \cite{Intriligator:2006dd,Intriligator:2007cp,Peskin:1997qi}, and then
introduces a deformation of this pure $SU(N_C)$ gauge theory by assuming the presence of one massive quark, then, upon integrating our this heavy quark, 
one obtains an effective superpotential of exactly the type used in smooth 
hybrid inflation \cite{Lazarides:1995vr,Lazarides:1996rk}. Therefore, the theory 
naturally embeds two mass scales: one, the strong-coupling scale, is generated dynamically, while 
the other is related to the heavy quark mass. Hence a salient feature of such SQCD-embedded inflation 
model lies in the existence of a UV completion of the theory. 
Although in its original form the framework of smooth hybrid inflation embedded in supergravity    
is not consistent with all observational constraints, this
can be corrected by considering a modified K\"ahler 
potential as shown in \cite{Rehman:2014rpa,urRehman:2006hu,Rehman:2012gd,Khalil:2012nd}.

In this work, we demonstrate that DM candidates can arise from the same SQCD sector. 
We first note that at the end of the smooth hybrid inflation, one field from the inflation 
system gets a vacuum expectation value (VEV) and therefore breaks the associated global 
symmetry of the SQCD sector yielding a number of Nambu-Goldstone bosons (NGBs), whose interactions with the SM are suppressed by 
the spontaneous symmetry breaking scale, which is large, being related to the scale of inflation. 
Nambu-Goldstone bosons as DM has already been studied in some other
 contexts, see
  for
   example, \cite{Berezinsky:1993fm,Boucenna:2014uma,Bhattacharya:2013kma,Garcia-Cely:2017oco,Ballesteros:2016euj,Frigerio:2011in,Rojas:2017sih,Lattanzi:2008ds,Gu:2010ys,Ametani:2015jla}.

We then assume a small deformation 
of the UV theory by providing small masses (small compared to one heavy 
quark mass already present in the construction with $N_f = N_c +1$ gauge theory) to the SQCD fermions; 
this generates the masses of NGBs through Dashen's formula
\cite{Veneziano:1983xe,Dominguez:1978qv}. Even with this deformation the NGBs are naturally 
stable and therefore serve as weakly interacting massive DM candidates; we will discuss this possibility in detail for various mass configurations.
In the non-degenerate case, we show that DM-DM interactions play a crucial role in the thermal freeze-out, and therefore 
in relic density, and also impose the spin-independent (SI) direct search constraints from the XENON 1T.

We will assume that visible matter is included in a supersymmetric sector that is well described by 
the minimal supersymmetric Standard Model (MSSM)
\cite{Haber:1984rc,Drees:2004jm,Wess:1992cp,Martin:1997ns,Nilles:1983ge} at low energies. In this case 
the lightest supersymmetric particle (LSP) also serves as DM candidate in this framework.
 The case where this LSP contributes an important share of the relic density has been explored in various 
publications\cite{Jungman:1995df,Ellis:2010kf,Steffen:2007sp,Olive:2004fp} within the MSSM.
 Here we consider an alternative scenario where the LSP relic abundance is small, while NGB-LSP interaction
 plays a crucial role in surviving the direct search constraints, particularly for degenerate NGB DM scenario.

The paper is organized as follows. In section \ref{sec:inflation.sqcd}, we discuss the basic SQCD framework, which leads to 
the smooth hybrid inflation. We point out the prediction of such smooth hybrid inflation model in view 
of PLANCK result\cite{Ade:2015xua}. Then in section \ref{sec:NGB-DM}, NGBs are identified as DM and a strategy 
for introducing DM masses are discussed. We indicate here on the superpotential 
that would be responsible for generating DM interaction with the SM particles. Parameter space scan 
for relic density and direct search constraints on the model  is elaborated in section \ref{sec:relic-direct} 
and we finally conclude in section \ref{sec:conclusion}.

%%%%%%%%%%%%%%%%%%%%%%%%%%%%%%%%%%%%%%%
\section{Smooth Hybrid Inflation in SQCD}
\label{sec:inflation.sqcd}
%%%%%%%%%%%%%%%%%%%%%%%%%%%%%%%%%%%%%%%%

We start with a brief introduction to the SQCD framework that leads to a smooth hybrid 
inflation as was proposed in \cite{Brax:2009yd}. We consider the existence of a strongly coupled supersymmetric 
$SU(N)$ gauge sector having $N_f$ flavors of quark superfields denoted by $Q_i$ and 
$\bar{Q_i}$ $(i=1,....,N_f)$ transforming as fundamental ($N$) and anti-fundamental ($\bar{N}$) 
representation of the gauge group $SU(N)$ respectively. This theory also has a global symmetry: 
$SU(N_f)_L\times SU(N_f)_R\times U(1)_B\times U(1)_R$, where the first $U(1)$ is proportional to 
the baryon number and the second one is related to the anomaly-free $R$-symmetry. We are particularly 
interested in $N_f = N$ case, where in the electric (or UV) theory, the following gauge invariant (but unnormalized) operators
can be constructed: $M_{ij} = {Q_i\bar{Q}_j}, ~ b = 
\epsilon_{i_1 i_2...i_N} \epsilon_{a_1 a_2...a_N} Q_{i_1}^{a_1}....Q_{i_N}^{a_N}$ and $\bar{b} 
= \epsilon_{i_1 i_2...i_N} \epsilon_{a_1 a_2...a_N} \bar{Q}_{i_1}^{a_1}....\bar{Q}_{i_N}^{a_N}$. 
Here $a_i$ correspond to the color indices and $i_j$ denote 
the flavor indices. 

Classically, in absence of any superpotential, these invariant 
operators are required to satisfy the gauge and flavor-invariant constraint ${\rm{det}} M - b\bar{b} = 0$.  As explained in 
\cite{Intriligator:2006dd,Intriligator:2007cp,Peskin:1997qi}, this constraint is modified by nonperturbative quantum contribution and becomes 
\begin{equation}
{\rm{det}} M - b\bar{b} = \Lambda^{2N}, 
\end{equation}
where $\Lambda$ is a dynamically-generated scale.
The corresponding quantum superpotential can be constructed by introducing a 
Lagrange multiplier field $X$, carrying $R$ charge of 2 units, and is given by  
\begin{equation}
W = X \Big({\rm{det}} M - b\bar{b} - \Lambda^{2N} \Big).
\label{eq:WN}
\end{equation}
The necessity of introducing $X$ follows from the fact that the expression of the 
quantum constraint does not carry any $R$ charge and the superpotential $W$ 
should have a $R$ charge of 2 units. 

With $N_f = N = 2$, it was shown in \cite{Dimopoulos:1997fv,Craig:2008tv} that such a superpotential results into 
a low energy effective superpotential that is very much similar to the one responsible 
for supersymmetric hybrid inflation. However, since in this case the predictions of the supersymmetric hybrid 
inflation are not in accordance with the results of WMAP and PLANCK, we use \cite{Brax:2009yd} instead $N_f = N = 4$, for which the corresponding effective 
superpotential still resembles the one in the smooth hybrid inflation scenario. In this case, the low energy 
(or IR) theory below the strong coupling scale $\Lambda_0$ of the $SU(N=4)$ gauge 
theory, can be described in terms of meson fields $T_{ij}$, baryon $B$ and antibaryon 
$\bar{B}$ superfields fields~\footnote{Here the $Q$ denote the $SU(4)$ quark super-fields.}
\begin{equation}
T_{ij} =\frac{Q_i\bar{Q}_j}{\Lambda_0}, ~~B = \frac{1}{\Lambda_0^3} \epsilon_{abcd} Q^a_1 Q^b_2 Q^c_3 Q^d_4,  
~~{\rm{and}}~~ \bar{B} =  \frac{1}{\Lambda_0^3} \epsilon_{ijkl} \epsilon_{abcd} \bar Q^a_1 \bar Q^b_2 \bar Q^c_3 \bar Q^d_4, 
\label{m-b}
\end{equation}
having the superpotential 
\begin{equation}
W = S \Big(\frac{{\rm{det}} T}{\Lambda_0^2} - B\bar{B} - \Lambda_{{\rm{eff}}}^{2} \Big).
\label{eq:W4}
\end{equation}
(the relation to Eq.(\ref{eq:WN}) is, $ T=M/\Lambda_0,~B=b/\Lambda_0^3$
where $\Lambda_0$ is the strong coupling scale of the $N_f=N=4$ theory; then $S$ can be identified with $\frac{X}{\Lambda_0^6}$ and $\Lambda_{\textrm{eff}}^2$ with $ \frac{\Lambda^8}{\Lambda_0^6}$)
% The connection between $\Lambda_0$ and
% $\Lambda_{\textrm{eff}}$ has been defined later (paragraph below Eq.(8)).]}
The effective mass scale $\Lambda_{{\rm{eff}}}$ can be interpreted in terms of 
holomorphic decoupling of one heavy flavor of quark (heavier than $\Lambda_0$) 
from a $SU(N)$ SQCD theory with $N_f = N + 1$ flavors as we discuss below.

%%%%%%%%%%%%%%%%%%%%%%%%%%
%%%%%%%%%%%%%%%%%%%%%%%
%%%%%%%%%%%%%%%%%

%%%%%%%%%%%%%%%%%%%%%%%%%%%%%%%%%%%%%%%%%
\subsection{Realization of the effective superpotential from $N_f = N + 1$ SQCD} 
\label{subsec:eff-superpot}
%%%%%%%%%%%%%%%%%%%%%%%%%%%%%%%%%%%%%%%%%

The low energy version of 
the supersymmetric $SU(N)$ gauge theory with $N_f = N + 1$ (where $N=4$) flavors is 
associated with mesons and baryons which are defined analogously to Eq. (\ref{m-b}), but 
due to the presence of an extra flavor ($i = 1, 2, ...5$), the baryons carry a free flavor 
index ($\hat{B}^i \propto \epsilon^{ijklm}\epsilon_{abcd} Q^a_j Q^b_k Q^c_l Q^d_m$) and the meson matrix ($\hat{T}_{i j}$) becomes correspondingly larger. 
Hence the baryons $\hat{B}^i$ and $\bar{\hat{B}}^i$ transform under the global group, 
$SU(N_f) \times SU(N_f)$, as $(N_f, 1)$ and $(1, \bar{N_f})$ respectively. 
Following Seiberg's \cite{Intriligator:2006dd,Intriligator:2007cp} prescription, the system can then be represented by the 
superpotential, 
\begin{equation}
\hat{W}_{m}=\hat{B^i}\hat{T_{ij}}{\bar{\hat B}}^j-\frac1{\Lambda_0^2} \textrm{det} \hat{T},\end{equation} 
where ${\Lambda}_0$ is the strong coupling scale of  $SU(N = 4)$ SQCD.
Note that the superpotential will have an $R$ 
charge 2 if we assign only the $\hat{T}_{N_f N_f}(=t) $ meson to carry $R=2$. 

Next we introduce a tree level quark mass term in the superpotential,
\begin{align}
\hat{W}_{N_f=N_c+1}=\hat{B}\hat{T}\bar{\hat{B}}-\frac1{\Lambda_0^2} \textrm{det} \hat{T}+{\Lambda}_0 \textrm{Tr}(\hat{m}\hat{T}).\textrm{ , } ~~{\rm{with}}~~
\hat{m} = {\rm{diag}}\{m_1,m_2,m_3,m_4, m_Q\}.
\label{eq:wm}
\end{align}
where we assume $ m_Q \gg m_{1,2,3,4}$.

Considering first the case $m_Q > \hat{\Lambda}_0$ and $m_{i=1,2,3,4} = 0$ (later we will discuss the effect of having nonzero $m_i  \ll m_Q$),
the $F$-flatness conditions for $\hat{T}_{i N_f}, ~\hat{T}_{N_f i}, \hat{B}^i, \bar{\hat{B}}^i$ (for $i < 5$) implies
\begin{equation}
\hat{B}=\begin{pmatrix}
  0 & B^{5}
 \end{pmatrix}
\textrm{, }\quad
\hat{\bar{B}}=
 \begin{pmatrix}
  0\\ 
  \bar{B}^{5}
 \end{pmatrix}, \quad
\textrm{ and  } \quad
\hat{T}=
\begin{pmatrix}
  T&0\\ 
  0& t
 \end{pmatrix},
\end{equation}
where we define the meson matrix ${T}_{ij} = \hat{T}_{ij}$ with $i, j = 1,2,3,4$ and $T_{55} = t$. 
Hence after integrating out the heavy $N_f$th (the 5th) flavor of quark, we are left with 
the following effective superpotential for $N_f = N = 4$ SQCD  
\begin{align}
W_{N_f=N_c}=t \Big(B_{5}\bar{B}_{5}-\frac{\textrm{det}T}{\Lambda_0^2}+m_Q\Lambda_0\Big).
\label{effective}
\end{align}
Comparing the above expression with Eq.(\ref{eq:W4}), we can now identify $B = \hat{B}^5, \bar{B} = 
\hat{\bar{B}}^5$ and $S = t$. Hence the effective mass parameter involved in Eq.(\ref{eq:W4}) 
is determined by the relation, $\Lambda_{\rm eff}^2 = m_Q \Lambda_0$ and the Lagrange 
multiplier field turns out to be proportional to the $N_f$th meson of the $N_f = N+1$ theory.

We now turn our attention to the superpotential in Eq.(\ref{eq:W4})  of $N_f = N (= 4) $ SQCD or 
equivalently to Eq.(\ref{effective}) and discuss the vacua of the theory. Different 
points on the quantum moduli space associated with this $N_f = N (= 4)$ SQCD 
theory exhibits different patterns of the chiral symmetry breaking \cite{Intriligator:2006dd,Intriligator:2007cp,Peskin:1997qi} . Here 
we are interested in the specific point on the quantum moduli space ($a ~la$ Eq.(\ref{eq:W4})) 
where $B = \bar{B} = 0$, and $T^{ij} = (\Lambda_0 \Lambda_{\rm{eff}})^{1/2} \delta_{ij}$, which is the global vacuum of the theory.  The corresponding chiral symmetry 
breaking pattern is then given by, 
\begin{equation}
SU(4)_L \times SU(4)_R \times U(1)_B \times U(1)_R \rightarrow SU(4)_V \times U(1)_B \times U(1)_R. 
\end{equation}
Hence along the direction $B=\bar{B}=0$ (the so-called meson branch of the theory), the superpotential reduces to
\begin{align}\label{infSup}
W_{\rm{Inf}}=S\Big(\frac{\chi^4}{\Lambda_0^2}-\Lambda_{\rm eff}^{2}\Big),
\end{align}
with det${T} = \chi^4$ and $ S = - t $; this superpotential is the same as the one used in the smooth hybrid inflationary 
scenario\cite{Lazarides:1995vr,Lazarides:1996rk}. 
The SQCD construction of the superpotential serves as a UV completed theory and also the scales 
associated are generated dynamically. Below we discuss in brief the inflationary predictions derived from this 
superpotential which can constrain the scales involved, $\Lambda_0, \Lambda_{\rm{eff}}$.   

%%%%%%%%%%%%%%%%%%%%%%%%%%%%%%%%%%%%%%%%%%%
\subsection{Inflationary predictions} 
\label{subsec:inflation}
%%%%%%%%%%%%%%%%%%%%%%%%%%%%%%%%%%%%%%%%%%%

The scalar potential obtained from Eq.(\ref{infSup}) is given by 
\begin{align}
V (\sigma, \phi)=\Big(\frac{\phi^4}{ 4\Lambda_0^2} - \Lambda_{\rm{eff}}^2 \Big)^2+\frac{\phi^6\sigma^2}{ \Lambda_0^4},
\end{align}
where the real, normalized fields are defined as $\phi =\sqrt{2}\chi$ and $\sigma = \sqrt{2}S$ 
(we use the same letters to denote the superfields and their scalar components). As 
was found in \cite{Lazarides:1995vr,Lazarides:1996rk}, the scalar potential has a local maximum at $\phi = 0$ for any value of the inflaton 
$\sigma$ and there are two symmetric valleys of minima denoted by $\langle \phi\rangle = \pm 
\Lambda_0 \Lambda_{\rm{eff}}/(\sqrt{3} 
\sigma)$. These valleys contain the global supersymmetric minimum
\begin{equation}\label{sumin}
\langle\phi\rangle=(2\Lambda_{\textrm{eff}} \Lambda_0)^{1/2},\textrm{  }\langle\sigma\rangle=0,
\end{equation}
which is consistent with our chosen point on the quantum moduli space given by 
det$T = \Lambda_0^2 \Lambda_{\rm{eff}}^2$. At the end of inflation  $ \sigma $ and $ \phi $ will roll down 
to this (global) minimum During inflation $\sigma^2\gg \Lambda_{\textrm{eff}}\Lambda_0$
and  $\phi$ is stabilized at the local minimum $\langle \phi\rangle = \pm 
\Lambda_0 \Lambda_{\rm{eff}}/(\sqrt{3} \sigma)$. The inflationary 
potential along this valley is given by
 \begin{equation}\label{effP}
V(\sigma)\simeq\Lambda_{\textrm{eff}}^4\Big(1-\frac{1}{54}\frac{\Lambda_{\textrm{eff}}^2\Lambda_0^2}{\sigma^4}\Big),
\end{equation}
for $\sigma^2\gg \Lambda_{\textrm{eff}}\Lambda_0$.

Within the slow-roll approximation, the amplitude of curvature perturbation $\Delta_R$, 
spectral index $n_s$, and the tensor to scalar ratio $r$ are given by
\begin{align}
\Delta_R^2=\frac{1}{24\pi^2M_P^4}\Big(\frac{V(\sigma)}{\epsilon}\Big),\\
n_s\simeq 1+2\eta-6\epsilon=1-\frac{5}{3N_e},\\
 r=16\epsilon=\frac{8(2\pi\Delta_R)^{2/5}}{27 N_e^2}
\end{align}
where $\epsilon$ and $\eta$ are the usual slow roll parameters. Assuming 
 $\langle\phi\rangle=M_{\textrm{GUT}}=2.86\times 10^{16}$ GeV, then $\Delta_R=2.2\times 10^{-9}$\cite{Ade:2015xua} implies 
 $\Lambda_0\simeq 4.3\times 10^{17} $ GeV and $\Lambda_{\textrm{eff}}\simeq1.8\times 10^{15}$ GeV. 
 When the number of e-folds is $N_e=57$, smooth hybrid inflation predicts $n_s\simeq  0.967$\cite{Senoguz:2003zw} and the very small value
 $r\simeq 3\times 10^{-6}$\cite{Senoguz:2003zw} that are in good agreement with Planck 2016 data\cite{Ade:2015xua}. 
   
  However, supergravity corrections to inflationary potential in Eq.(\ref{effP}) have important contributions.
  With minimal K\"ahler potential, previously obtained values of $n_s$ and $r$ change to $0.99$ and $1\times 10^{-6}$
  respectively\cite{Senoguz:2003zw}. Particularly the value of $n_s\sim 0.99$ is in tension with  observations \cite{Ade:2015xua}.
   To circumvent the problem one may use non-minimal K\"ahler potential as suggested in 
   \cite{Khalil:2012nd} to bring back the value of $n_s$ within the desired range. On the other hand, to increase $r$ to a detectable limit,
  further modification in K\"ahler potential may be required, as shown in \cite{Rehman:2014rpa,Rehman:2012gd}.

%%%%%%%%%%%%%%%%%%%%%%%%%%%%%%%%%%%%%%%%%
\section{NGB as dark matter in SQCD}
\label{sec:NGB-DM}
%%%%%%%%%%%%%%%%%%%%%%%%%%%%%%%%%%%%%%%%%

Once the inflation ends, the inflaton system slowly relaxes into its supersymmetric ground state 
specified by Eq.(\ref{sumin}). This however spontaneously breaks the associated flavor 
symmetry from $SU(4)_L \times SU(4)_R $ to the diagonal $SU(4)_V$ subgroup. This would 
generate fifteen Nambu-Goldstone bosons (NBGs) those are the lightest excitations in the model. 
The Lagrangian so far considered has only the usual derivative couplings among the NGBs, which are 
suppressed by inverse powers of $ \langle\chi\rangle$; in particular, they have no interactions with the
SM sector, and they are  stable over cosmological time scale. In the following we will introduce additional
interactions that will modify this picture.

At the global minimum we can write 
\begin{align}\label{Tvev} 
T =\chi\exp\Big(\frac{iG_S^a\lambda^a}{\langle\chi\rangle}\Big),
\end{align}
where $ G^a,\, (a=1,\ldots,15) $ denote the NGBs, and $ \lambda_a$ are the generators of $ SU(4) $.
This expression also identifies $ \langle\chi\rangle$ as the equivalent of the pion decay constant. 
Up to this point the NGBs are massless, which
can be traced back to the choice of $ m_{i=1,2,3,4}=0$ in Eq. (\ref{eq:wm}). If we relax this assumption (while maintianing $ m_i \ll m_Q$) the chiral symmetry is broken (explicitly) and, as a consequence, 
the NGBs acquire a mass that can be calculated using 
Dashen's formula\cite{Veneziano:1983xe,Dominguez:1978qv}:
\begin{align}
\langle\chi\rangle^2 (m_{G_S}^a )^2&=\langle0|[\tilde{\mathcal{Q}_a},[\tilde{\mathcal{Q}_a},H]]|0\rangle\qquad (\mbox{no~summation~over~}a)
\nonumber \\
&=\bar{\psi}\Big[\frac{\lambda_a}{2},\Big[\frac{\lambda_a}{2},m_{\textrm{diag}}\Big]_{+}\Big]_{+}\psi\,; \qquad m_{\textrm{diag}} = \textrm{diag}
\{m_1,\,m_2,\,m_3,\,m_4\}
\label{eq:mdiag}
\end{align}
where, as noted above, $ \langle\chi\rangle$ corresponds to the decay constant,  the ``$+$" subscript denotes the anticommutation, and
$\tilde{\mathcal{Q}_a}= \frac12\int d^3x\psi^{\dagger}\gamma_5\lambda_a\psi$ are the $SU(4)_A$ axial charges 
with the quark state $\psi = (Q_1, Q_2, Q_3, Q_4)^T$; $\lambda_a$ are the $SU(4)$ generators normalized such that $ {\rm Tr} 
[\lambda_a \lambda_b] = \frac{ \delta_{ab}}{2} $ ($a,b,\ldots$ denote  generator indices with $a=1,2,....,15$). The details of the mass spectrum of these pseduo-NGBs (pNGB's) are provided 
in Appendix I. 

The pNGB spectrum is determined by the light-mass hierarchy. Among several possibilities, we will concentrate on the following two cases:
\begin{itemize}
\item The simplest choice is to take $m_{1,2,3,4}=m$ in $m_{\textrm{diag}}$ (see Eq.(\ref{eq:wm})).
 In this case all the fifteen pNGBs will be degenerate, having mass $m_{G_S}^2=2m \Lambda^3/{\langle\chi\rangle^2}$, where $\Lambda^3 =\langle \bar{Q}_iQ_i \rangle$.
  
\item A split spectrum can be generated if one assumes $m_1=m_2=m_3=m_\gamma$
 and $m_4\gg m_{\gamma}$. Then we find three different sets of pNGB with different masses: {\it(i)} eight pNGBs will have mass $m_A^2=2m_\gamma  {\Lambda^3}/{\langle\chi\rangle^2}$; {\it(ii)} six pNGBs with mass
 $m_B^2= (m_4+m_\gamma) {\Lambda^3}/{\langle\chi\rangle^2}$; and {\it(iii)} 
 one pNGB with mass $m^2_C=\Big(3m_4+m_\gamma \Big)  {\Lambda^3}/{2\langle\chi\rangle^2}$ (see 
 Appendix I for details).
\end{itemize} 

We now turn to the discussion of the interactions of these pNGBs with the SM in this set-up. As noted in the introduction, we assume that at low energies the SM is contained in the MSSM. In this case, the $S$ field (being neutral) serves as a mediator between the MSSM and SQCD sectors which then leads to the following modified superpotential
\begin{align}\label{totP}
W_{\rm T}=S\Big(\frac{\textrm{det} T}{\Lambda_{0}^2}-\Lambda_{\rm eff}^2\Big)+\kappa_1 S\Big\{\textrm{Tr}(T^2)-\frac{(\textrm{Tr }T)^2}{N_f} \Big\}+\kappa_2 S H_u H_d,
\end{align}
where $H_u$ 
and $H_d$ are the two  Higgs (superfield) doublets in MSSM, and 
$ \kappa_{1,2}$ are phenomenological constants that can be taken positive. 

The  terms within the curly brackets are phenomenologically motivated additions. 
Note that while the first (as in Eq. \ref{infSup}) and last terms in $ W_{\rm Inf}$ respect the full $SU(N_f)_L\times SU(N_f)_R\times U(1)_B\times U(1)_R$ chiral symmetry, the middle term only respects the diagonal subgroup \cite{Giveon:2007ef}. The 
inclusion of such terms hence naturally lead to additional interactions of pNGBs. It is important to note  that during inflation $T$ is proportional to the unit matrix, so the term proportional to $ \kappa_1$  vanishes and  does not affect the smooth hybrid inflation scenario described before. Below we will see how incorporation of such 
terms can lead to the desired DM properties. 

The  term in $W_{\rm T}$ proportional to $ \kappa_2 $ provides a connection between the SQCD sector and the minimal supersymmetric Standard model (MSSM). Inclusion of this term in the superpotential will have a significant effect 
on reheating after inflation\cite{Brax:2009yd}; it can address the so-called $\mu$-problem  in the MSSM\cite{Dvali:1996cu,Kim:1983dt,Dvali:1997uq}, provided $S$ acquires a small, $\sim\mathcal{O}(\mbox{TeV})$, vacuum expectation value. We will see that it also provides a useful annihilation channel for DM. 

The linearity on $S$ in these new contributions to 
the superpotential is motivated from $R$ symmetry point of view. We note that any dimensionless coupling multiplying the 
first term in $W_{\rm Inf}$ can be absorbed in a redefinition of $ \Lambda_{0, \rm eff}$; in contrast, the 
couplings $ \kappa_{1,2}$ (assumed real) are physical and, as we show below, are constrained by observations 
such as the dark matter relic abundance and direct detection limits.

The scalar potential can now be obtained from $V_{\textrm{scalar}}= |{\partial W_{\rm T}}/{\partial S} |^2+ |{\partial W_{\rm T}}/{\partial T}|^2$. 
At the supersymmetric minimum, it is given by
\begin{align}\label{scalar}
V_{\textrm{s}} =\frac{\kappa_1^2}{4}\sum_{a,b}(G_S^a)^2(G_S^b)^2 +\kappa_2^2|H_u|^2|H_d|^2-\frac{\kappa_1\kappa_2}{2}\sum_a(G_S^a)^2
H_uH_d+\textrm{h.c.} + \cdots
\end{align}
In obtaining this, we expanded $T$ in powers of the NGBs:
\begin{align}
T =\chi\Big\{1+\frac{iG_S^a\lambda^a}{\langle\chi\rangle}-\frac{G_S^a G_S^b \lambda^a\lambda^b}{2\langle\chi\rangle^2}+.... \Big\},
\end{align}
and $ \chi$ is developed around its expectation value: $\chi = \langle\chi\rangle + \cdots$.

Note that the first term in $V_{\textrm{s}}$ is of interest only when the pNGBs are not degenerate 
as otherwise it would not contribute to number changing process. In such a case with non-degenerate 
pNGBs, the heavier $G$s can annihilate into the lighter ones. Hence in such a situation, $\kappa_1$ 
can also play a significant role in our DM phenomenology along with $\kappa_1 \kappa_2/2$. In fact,
we will show that the annihilation of the heavier ones to the lighter components will aid in freeze-out 
of the heavier component. This helps in evading the direct search bounds as the coupling $\kappa_1$ 
alone will not contribute to direct search cross section, thereby allowing a larger parameter space viable 
to our DM scenario. It is important to note here that even if we assume that the masses of the heavier 
pNGBs are very large, their annihilation cross-sections to the SM will be small enough for an early 
freeze-out, which leads to an unacceptably large  relic abundance unless a large enough $\kappa_1$ 
allows them to annihilate to lighter ones. Note that the interactions among the $G$ generated by  
Tr$(\partial_\mu T^\dagger\, \partial^\mu T)$ are negligible since they are suppressed by powers 
of $\langle\chi\rangle$.

%\begin{align}
%\partial_\mu T=\partial\chi+\chi\frac{i\partial G_S^a\lambda_a}{\langle\chi\rangle-
%\chi\frac{(G_s^b\partialG_s^a+G_s^a\partialG_s^b)\lambda_a\lambda_b}{2\langle\chi\rangle}^2}\\
%\end{align}

The interaction of the pNGBs with the MSSM sector (the last term in Eq.(\ref{scalar})),  is Higgs-portal like:
\begin{align}\label{int1}
V_{\textrm{Int}}  = -\lambda\sum^{15}_{a=1}  (G_S^a)^2(H_u^+H_d^- -H_u^0H_d^0)\,, ~~{\rm{where}}
~~\qquad \lambda=\frac{\kappa_1\kappa_2}{2}.
\end{align}
In terms of the physical mass eigenstates, the two Higgs doublets in MSSM can be written as follows\cite{Haber:1984rc,Drees:2004jm,Wess:1992cp,Martin:1997ns}: 
\begin{equation}\label{H1}
H_u=
\begin{bmatrix}
H_u^-\\
H_u^0
\end{bmatrix}=\frac{1}{\sqrt{2}}
\begin{bmatrix}
    \sqrt{2}(H^- \sin\beta-X^- \cos\beta) \\
    v_u+(H \cos\alpha-h \sin\alpha)+i (A \sin\beta+X^0 \cos\beta) 
\end{bmatrix},
\end{equation}
\begin{equation}\label{H2}
H_d=
\begin{bmatrix}
H_d^0\\
H_d^+
\end{bmatrix}=\frac{1}{\sqrt{2}}
\begin{bmatrix}
    v_d+(H \sin\alpha+h \cos\alpha)+i (A \cos\beta-X^0 \sin\beta)  \\
     \sqrt{2}(H^+ \cos\beta+X^+ \sin\beta)
\end{bmatrix},
\end{equation}
where $h$ and $H$ denote the light and heavy CP-even eigenstates respectively; $ H^\pm$ and $A$  are 
the charged and CP-odd physical scalars respectively, and $ X^{0,\pm}$ are the would-be Goldstone bosons. As usual, 
$h$ plays the role of the SM Higgs, and the vacuum expectation values of $H_u^0, ~H_d^0$ (denoted by $v_u$ and $v_d$ 
respectively) are related by 
\begin{align}
\tan\beta = \frac{v_u}{v_d}\,, \qquad v = \sqrt{v_u^2 + v_d^2} \simeq 246 \textrm{ GeV}.
\end{align}
The other mixing angle $\alpha$ appears as a result of the diagonalization of the CP-even Higgs 
mass-squared matrix (in the $H_u^0- H_d^0$ basis) leading to the physical Higgses, 
$h$ and $H$. The mixing angle $\alpha$ can be expressed in terms of $\beta$ and the pseudoscalar $A$ mass as 
\begin{align}
\tan 2\alpha&=\tan 2\beta\, \frac{M_A^2+M_Z^2}{M_A^2-M_Z^2}. 
\label{eqtan1}
\end{align}

We can now write the coupling of the pNBGs with the SM Higgs from Eq.(\ref{int1}) as follows:
\begin{equation}
  V \supset - \left( \lambda^{'} \, h^2 +  \lambda^{''} \, h v_d \right) \sum_a (G^a_S)^2 \,,
  \label{eq:int}
  \end{equation}
  where 
\begin{equation}
  \lambda^{'}=\frac{1}{2}\lambda \sin\alpha \cos\alpha,\quad
  \lambda^{''}=\frac{1}{2}\lambda \left( \sin\alpha - \frac{v_u}{v_d} \cos\alpha \right)=\frac{1}{2}\lambda  \, \cos\alpha(\tan\alpha - \tan \beta).
  \label{eq:lambda}
\end{equation}
On the other hand, the couplings of the SM Higgs $h$ to the vector fields and fermions 
are given by
\begin{equation} 
hWW : \frac{2 m_W^2}{v} \sin (\beta-\alpha), ~~
hZZ : \frac{2 m_Z^2}{v} \sin (\beta-\alpha), ~~
hf\bar{f} : \frac{m_f}v \, \frac{\sin{\alpha}}{ \cos \beta}.
\label{smc}
\end{equation}
In view of Eq. (\ref{eqtan1}), it can be noted that in the large pseudoscalar Higgs 
mass limit $M_A \gg M_Z$, $\tan 2\beta \simeq \tan 2\alpha$, that has two solutions. 
One possibility is  $ \alpha \simeq \beta $, in which case  couplings 
of the SM Higgs with $W$ and $Z$ vanish (see Eq.(\ref{smc})) and $\lambda^{''} \sim \cos 
\alpha(\tan \beta -\tan \alpha) \to 0$ (see Eq.(\ref{eq:lambda})), and hence is of limited interest.
\begin{figure}[thb]
$$
 \includegraphics[height=6.5cm,width=8cm]{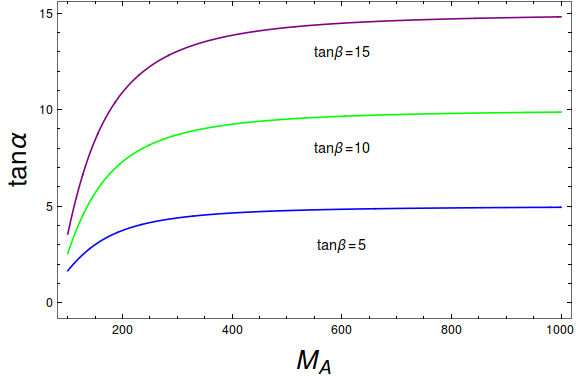}
$$
\caption{$\tan \alpha$ versus $M_A$ where
%for (left) $\alpha=\beta$ and (right)
 $\alpha=\frac{\pi}{2}+\beta$.}
\label{fig:ma-tb}
\end{figure}
There is, however, a second solution: 
$\alpha \simeq\beta+\pi/2 $, so that $ \tan \beta \simeq -\cot \alpha$ (see Fig.\ref{fig:ma-tb}) 
in which case the lightest CP-even scalar $h$ will be SM-like and Eq.(\ref{eq:int}) closely resembles 
a Higgs-portal coupling\cite{Guo:2010hq,McDonald:1993ex,Biswas:2011td}. 
In the following we 
will continue to assume $ \alpha = \beta + \pi/2$. It is easy to show that with such a 
choice, $\lambda', \lambda^{''}$ and $\lambda$ are related with $\beta$ as (using Eq.(\ref{eq:int}) 
\begin{align}
\frac{\lambda^{''}}\lambda = \frac{1}{2\cos\beta}\,, \quad
\frac{\lambda^{\prime}}{\lambda}= -\frac{1}{4}\sin2\beta\,; \qquad (\alpha = \beta + \pi/2).
\end{align}

\begin{figure}[h]
 \includegraphics[height=4.5cm]{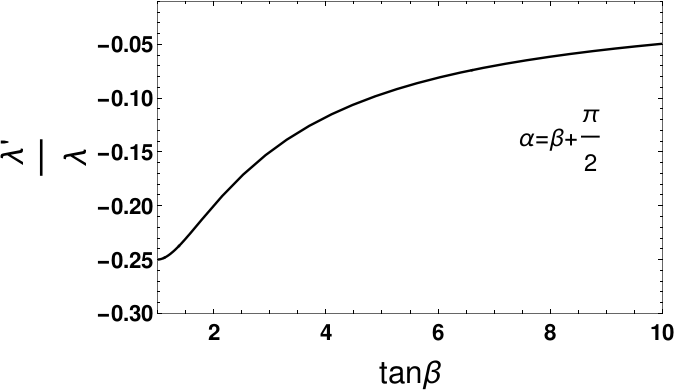}
 \includegraphics[height=4.6cm]{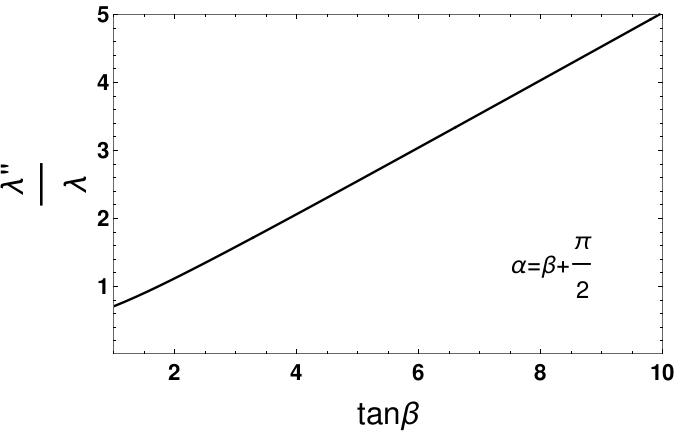}
\caption{ Variation of $\lambda^{\prime}$ and $\lambda^{''}$ ( scaled by $\lambda$ ) with $\tan\beta$. 
We assume $\alpha = \beta+ \pi/2$ for both the cases.}
\label{fig:lamb-beta}
\end{figure}

These couplings are plotted in Fig.\ref{fig:lamb-beta} as functions of $ \tan\beta $ (note that $ v_d$ 
is a monotonically decreasing function of $ \tan\beta$). 

%it will be useful below to, as shown in Fig.~\ref{fig:beta}  
%\begin{figure}[thb]
%$$
%% \includegraphics[height=6.5cm]{tb-vd}
%$$
%\caption{Plot of $v_d$ as a function of $ \tan\beta$.}
%\label{fig:beta}
%\end{figure}
 
%%%%%%%%%%%%%%%%%%%%%%%%%%%%%%%%%%%%%%%%
\section{Relic density and direct search of pNBG DM}
\label{sec:relic-direct}
 %%%%%%%%%%%%%%%%%%%%%%%%%%%%%%%%%%%%%%%%

We now turn to the determination of the regions of parameter space where our DM candidates, the pNGB's, 
satisfy the relic density and direct detection constraints.  The pNBGs interact with the SM through the Higgs 
portal as described in Eq.(~\ref{eq:int}), so the behavior of the pNBG sector is similar to that of a 
scalar singlet Higgs portal. In view of our consideration, $\alpha = \beta + \pi/2$, the parameters for this 
sector are then effectively 
\begin{align}
m_{G_s},\lambda, \tan \beta.
\end{align}
\noindent The other factor which determines the relic density of pNGB's is their mass spectrum; we will show that 
depending on the choice of mass parameters employed in the Dashen formula, we can have several 
phenomenologically viable situations where one or more of the $G$ contribute to the relic density.

We hasten to note, however, that this model also contains an additional particle in the dark sector: 
the lightest supersymmetric particle (LSP), that we assume to be the lightest neutralino ($\chi_0$), 
which is stable due to $R$ parity conservation. The MSSM neutralino $\chi_0$  has been studied 
as a DM candidate in various contexts\cite{Jungman:1995df,Ellis:2010kf,Steffen:2007sp,Olive:2004fp}, and can annihilate to SM and supersymmetric particles through 
many different but known channels, depending on the composition of Bino-Wino-Higgsino admixture. In this 
work we will concentrate on the case where the pNBGs dominate the DM abundance. We still
must incorporate neutralino and MSSM phenomenology to some extent as the pNBGs interact with the 
neutralino DM through the Higgs portal coupling. This is because, there is a Wino-Higgsino-Higgs ($\tilde{W}-\tilde{H}_{u/d}-H_{u/d}$) or Bino-Higgsino-Higgs ($\tilde{B}-\tilde{H}_{u/d}-H_{u/d}$)
coupling in the MSSM through which the pNBGs can annihilate into a pair of neutralinos (or vice versa, depending on the mass hierarchy of the pNGBs and LSP). 
The strength of the pNBG-LSP interaction of course depends on the composition of neutralino and we consider two different situations of 
phenomenological interest: {\it (i)} when the pNBG-LSP interactions can be completely neglected and the DM relic density is solely composed 
of pNBGs, and {\it (ii)} when the pNBG-LSP interaction is weak but non-vanishing.

In general, the coupled Boltzmann equations that determine the relic density for our two-component DM model 
(considering the presence of a single pNGB having mass $m_{G_s}$ and LSP with mass $m_{\chi}$) can be 
written as follows:
 \begin{align}
\nonumber
 \dot{n_{G_S}}+3Hn_{G_S}&=-\langle \sigma v \rangle_{G_SG_S\to SM} (n_{G_S}^2-{n_{G_S}^{eq}}^2) - \langle \sigma v \rangle_{G_SG_S \to \chi \chi} (n_{G_S}^2-n_{\chi}^2),\\ 
  \dot{n_{\chi}}+3Hn_{\chi}&=-\langle \sigma v \rangle_{\chi \chi \to SM} (n_{\chi}^2-{n_{\chi}^{eq}}^2) + \langle \sigma v
  \rangle_{G_SG_S \to \chi \chi} (n_{G_S}^2-n_{\chi}^2),
  \label{eq:begin{align}}
 \end{align}
 where we assume  $m_{G_s} > m_{\chi}$. Here $ n_{G_S}$ and $ n_\chi$ denote the number density for the pNBG and LSP, respectively. Corresponding equilibrium distributions are given by 
\begin{equation}
n_{i}^{eq} =  \int \frac{\xi_{i} d^3 p}{(2 \pi)^3 }  \tilde{f}_{i}^{eq},    \hspace{.4 cm} {\rm with\;} \hspace{.2 cm} \tilde{f}_{i}^{eq}  = \frac{1}{e^{E/k_B T}-1}\;.
\label{eq:no-density}
\end{equation}
where the index represents either of the DM species $i=\{G_s, \chi\}$ and $\xi_i$ denotes the degrees of freedom for the corresponding DM species, and we assume zero chemical potential for all particle species.
In principle annihilations of the pNBGs occur to both SM and supersymmetric particles. However, assuming that the supersymmetric particles
are heavier (as searches at LHC have not been able to find them yet), the dominant annihilation of the pNBGs occurs to SM particles. After freeze out, 
relic density of the DM system is described by 
\begin{align}
\Omega = \Omega_{G_S} +\Omega_{\chi},
\end{align}
where the individual densities  are determined by the freeze-out conditions of the respective DM components; which, in turn, are governed by the 
annihilation of these DM components to the SM, and as well as by the interactions amongst themselves. 

One can clearly see from Eq.~(\ref{eq:begin{align}}) that the Boltzmann equations for the two dark sector components, LSP and pNGBs, are coupled due to the presence of the
terms containing $\langle \sigma v \rangle_{G_SG_S \to \chi \chi}$; when this is  of the same order as  $\langle \sigma v \rangle_{G_SG_S\to SM}$
the individual abundances will differ significantly from  those obtained when $\langle \sigma v \rangle_{G_SG_S \to \chi \chi} \simeq 0 $. For example,
consider the  case where the LSP is dominated by the wino component. Then $\langle \sigma v \rangle_{\chi \chi \to SM}$ is significant because of the
large coupling of the wino to the $Z$, and from the co-annihilation channel involving the lightest chargino. Because of this large cross section
we expect that the LSP relic density will be small: $ \Omega_\chi \ll 0.1 $ and the relic density will be composed almost solely of pNBGs:
$ \Omega_{G_S}h^2 \sim \Omega h^2 \sim 0.1$. In the following we will consider separately the cases where the LSP-pNBG interactions 
are negligible and when they are significant.

As stated earlier, our scenario allows for 15 pNBGs which can be degenerate. Hence their 
total contribution to $ \Omega $ will be 15 times that of a single boson. However, the
degeneracy of the pNBGs follow from making the simplifying assumption $m_1=m_2=m_3=m_4$ in Eq.(\ref{eq:mdiag})(see discussions of Dashen's formula in Appendix I). Other choices generate  different patterns of the dark sector mass hierarchy, which in turn govern the phenomenology of the pNBG as DM candidates. 
  
  %%%%%%%%%%%%%%%%%%%%%%%%%%%%%%%%%%%%%%
 \subsection{Negligible pNBG-LSP interaction limit}
 %%%%%%%%%%%%%%%%%%%%%%%%%%%%%%%%%%%%%%

\subsubsection{Degenerate pNBG DM}
%%%%%%%%%%%%%%%%%%%%%%%%%%%%%%%%%%%%%
The simplest case we consider is that of completely degenerate pNGBs (which corresponds to $m_1=m_2=m_3=m_4$); in this case the 15 $G_S$
contribute to DM relic density equally. In the absence of pNGB-neutralino interactions the individual abundance by the single-component Boltzmann equation
\begin{align}
\dot{n}_{G_S}+3Hn_{G_S}=-\langle \sigma v \rangle_{G_SG_S\to SM} (n_{G_S}^2-n_{eq}^2).
\label{eq:case-a}
\end{align}
where the interactions between different pNBGs are also ignored because of the degeneracy. Therefore, total relic abundance can be obtained by adding all of the single component contributions. Now the relic density for a single component pNGB is given by\cite{Kolb:1990vq}
\begin{align}
 \Omega_{G_S}^1=\frac{m_{G_S}n_{G_S}^{x\rightarrow\infty}}{\rho_{\textrm{c}}}~\textrm{GeV}^{-2},
\label{eq:omS}
\end{align}
where $\rho_c=1.05\times 10^{-5} h^2 \frac{\textrm{GeV}}{c^2}\textrm{cm}^{-3}$\cite{Patrignani:2016xqp} is the critical density of the universe. Eq.(\ref{eq:omS}) can be
translated to 
\begin{align}
 \Omega_{G_S}^1h^2=\frac{8.51\times 10^{-12} x_f}{\langle\sigma v\rangle}~\textrm{GeV}^{-2},
\end{align}
where $x_f=\frac{m_{G_S}}{T_f}$ with the freeze-out temperature $T_f$. Next if one considers\footnote{We show in Appendix III that actual numerical solution 
to Boltzman equation, matches to approximate analytical solution for such values of $x_f$.} $x_f=22$, the total relic density turns out to be
\begin{equation}
 \Omega h^2 = 15~\Omega_{G_S}^1h^2\simeq15\times \left[\frac{2.0 \times 10^{-10}~{\rm GeV^{-2}}}{\langle \sigma v\rangle_{G_SG_S \to SM}} \right]
 \label{eq:omA}
 \end{equation}
 
\begin{figure}[thb!]
$$
 \includegraphics[height=6.4cm]{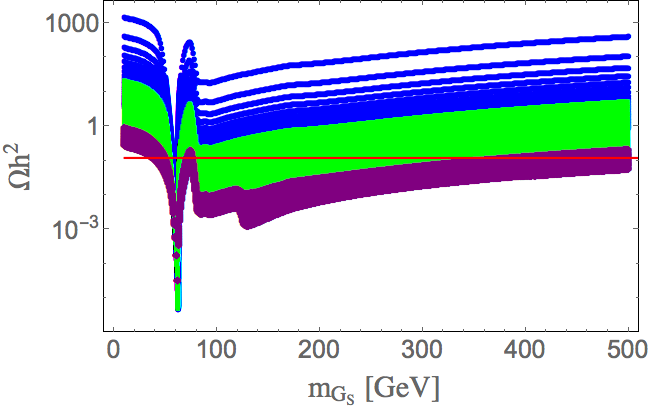}
$$
\caption{DM relic density as a function of the DM mass for the case of 15 degenerate pNBGs with $ \tan\beta=10$. 
The coupling $ \lambda$ is varied between $0.01-0.25$ (blue), $0.25-0.5$ (green) and $0.5-1.0$ (purple); 
The horizontal band shows the correct density by PLANCK. For the scan, we chose: $\alpha=\beta+\pi/2$.}
\label{fig:Omega1}
\end{figure}
 
As mentioned earlier, the annihilation of pNBGs to SM states is controlled by two couplings 
$\lambda^{'}$ and $\lambda^{''}$ through Eq.(\ref{eq:int}). This situation is similar
to the usual Higgs portal coupling with a singlet scalar, stabilized  under a $Z_2$ symmetry\cite{Guo:2010hq,McDonald:1993ex,Biswas:2011td}
, with the important difference that the model we consider has two independent couplings, determined by $ \lambda $
and the angles $ \alpha $ and $\beta $ (cf. Eq. \ref{eq:lambda}). In the limit where the psudoscalar mass is large, 
which we assume, the number of parameters is reduced by one because of the relation $ \tan 2\alpha = \tan 2 \beta $. 
This results into the phenomenologically acceptable relation involving $\alpha$ and $\beta$ as 
$\alpha = \beta + \pi/2$.

The important cross sections contributing to Eq.~(\ref{eq:omA}) are: 
\begin{eqnarray}
(\sigma v)_{G_s G_s \rightarrow f \overline f}&=&\frac{m_f^2}{\pi v^2} \frac{\lambda^{\prime\prime 2}v_d^2}{(s-m_h^2)^2+m_h^2 \Gamma_h^2}
\Big(1-\frac{4m_f^2}{s}\Big)^\frac{3}{2},\nonumber\\
{(\sigma v)}_{G_s G_s \rightarrow W^+ W^-}&=&\frac{\lambda^{\prime\prime 2}v_d^2}{2\pi v^2}  \frac{s}{(s-m_h^2)^2+m_h^2 \Gamma_h^2}(1+\frac{12m_W^4}{s^2}-
\frac{4m_W^2}{s})\Big(1-\frac{4m_W^2}{s}\Big)^\frac{1}{2},\nonumber\\
{(\sigma v)}_{G_s G_s\rightarrow Z Z}&=&\frac{\lambda^{\prime\prime 2}v_d^2}{4\pi v^2 } \frac{s}{(s-m_h^2)^2+m_h^2 \Gamma_h^2}
(1+\frac{12m_Z^4}{s^2}-\frac{4m_Z^2}{s})\Big(1-\frac{4m_Z^2}{s}\Big)^\frac{1}{2},\nonumber\\
{(\sigma v)}_{G_s G_s \rightarrow h h}&=&\frac{1}{16\pi s}\Big[4\lambda^\prime+\frac{6\lambda^{\prime\prime}v_dm_h^2}{v(s-m_h^2)}
-\frac{16\lambda^{\prime\prime 2}v_d^2}{(s-2m_h^2)}\Big]^2 \Big(1-\frac{4m_h^2}{s}\Big)^\frac{1}{2}.
\label{eq:annihilationSM}
\end{eqnarray}

The relic density $ \Omega $ will be a function of the pNBG mass $ m_{G_s} $, the coupling $ \lambda $ and the angles $ \alpha$, $\beta$. 
In Fig.~\ref{fig:Omega1} we evaluate $\Omega $ as a function of $ m_{G_s}$ for $ 0.05<\lambda<1.0$ and for $ \tan\beta=10 $ 
when $ \alpha = \beta + \pi/2 $; the evaluation is obtained using {\tt MicrOmegas}\cite{Belanger:2014vza}. Note here, that the dominant annihilation of pNBG DM 
comes through $ \lambda^{''} $. But the change in $ \lambda^{''} $ due to change in $\tan\beta$ is neatly balanced by the change in $v_d$ accompanying $\lambda^{''}$ in all vertices making the relic density invariant under $\tan\beta$. The results exhibit the usual 
resonant effect when $ m_{G_s} \sim m_h/2 \sim 62.5 $ GeV. We can also see that as $ \lambda $ increases 
$ \Omega $ drops, a consequence of having larger cross sections.

\begin{figure}[thb!]
$$
 \includegraphics[height=6.4cm]{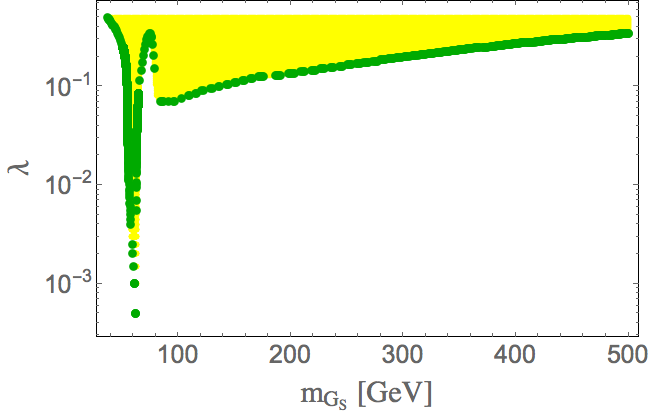}
$$
\caption{
Regions of the $ m_{G_S}-\lambda $ plane allowed by the relic density constraint by PLANCK\cite{Ade:2015xua} when all 15 pNBGs are degenerate is shown in green. Under abundant region is shown in yellow. We choose $\tan\beta = 10$ with $\alpha=\beta+\pi/2$. 
}
\label{fig:M1-K1}
\end{figure}

% \begin{figure}[thb!]
%$$
%% \includegraphics[height=5.5cm]{Omega-M-tb10-1.png}
%% \includegraphics[height=5.5cm]{Omega-M-tb5-1.png}
%$$
%$$
%% \includegraphics[height=5.5cm]{Omega-M-tb25-1.png}
%$$
%\caption{ 
%Relic DM density as a function of the DM mass \jw{for the case of a single lightest pNBG, and when $ \tan\beta=10,5$ and $2.5$ (upper left, upper right and bottom, respectively). The coupling $ \lambda$ is varied between $0.05-0.25$ (blue), $0.25-0.5$ (green) and $0.5-2.0$ (purple); in all the cases we chose: $\alpha=\beta+\pi/2$.}}
%\label{fig:Omega2}
%\end{figure}

%\begin{figure}[thb!]
%$$
%% \includegraphics[height=5.5cm]{M1-l-1.png}
%% \includegraphics[height=5.5cm]{M1-l-tb10-1.png}
%$$
%\caption{Regions of the $ m_{G_S}-\lambda $ plane allowed by the relic density constraint in a hypothetical situation when 
%{\em one} pNBG contributes as DM, for $\tan\beta = 2.5,\,5$ (left pane, orange and green points respectively) and $ \tan\beta =10$ 
%(right pane, blue points). In all the cases we choose: $\alpha=\beta+\pi/2$.}
%\label{fig:M1-K1-2}
%\end{figure}

The allowed region in the $ m_{G_S}-\lambda $ plane by the DM relic density constraint
is presented shown in Fig. \ref{fig:M1-K1} for $ \tan\beta=10$ when all 15 pNBGs are degenerate. The allowed parameter space is  similar to that of a Higgs portal scalar singlet DM. The difference is mainly due to our having 15 particles: we requires larger cross section, corresponding to a larger value of $ \lambda $, to compensate for the factor of 15 in Eq. \ref{eq:omA}. 

%\jwn{I think this discussion belongs to the next subsection}\textcolor{blue}{[Ans: You are right.]}
%When compared to the hypothetical case of a single component pNBG DM, 
%it is apparent that the correct relic density is obtained for smaller values of $ \lambda $. This is simple to understand: 
%when there are 15 degenerate NBGs the contribution from {\em each} to the relic density has to be 15 times smaller, which requires a larger 
%(individual) cross section, and corresponds to a larger value of $ \lambda $. But such a situation (fourteen degenerate and one non-degenerate DM) 
%is difficult to achieve in this pNBG framework from Dashen formula. Secondly, even if we imagine such 
%a hypothetical case, fourteen degenerate pNBGs heavy or light, are all stable due to $(G_s^a)^2(G_s^b)^2$ type of interactions and will contribute to 
%DM relic density. In both the extreme cases of being very heavy or light, they will have small annihilation cross-section and large relic density 
%(see for example, the behaviour in Fig.~\ref{fig:Omega1}) which will over close PLANCK\cite{Ade:2015xua} relic density bound and make the scenario phenomenologically unviable. Hence, having an effective single-component pNBG DM is not possible here. 

\begin{figure}[thb!]
$$
\includegraphics[height=6.4cm]{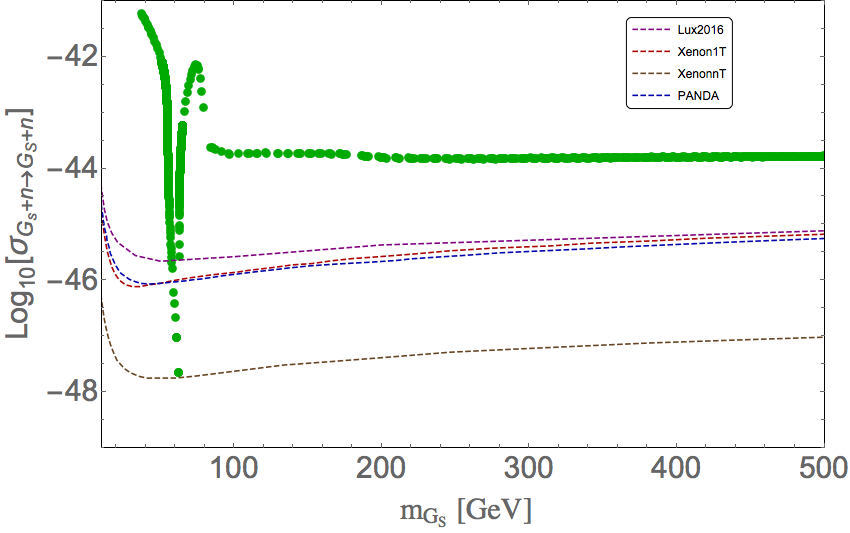}
$$
\caption{Spin-independent DM-nucleon cross section as a function of the DM mass, for $ \tan\beta = 10$ for the case of 15 degenerate pNBG DM. Bounds from LUX 2016
\cite{Akerib:2016vxi}, Xenon1T\cite{Aprile:2017iyp} and PANDA X\cite{Cui:2017nnn} are shown with the expected sensitivity from XENONnT\cite{Aprile:2015uzo}.}
\label{fig:DD1}
\end{figure}

The non-observation of DM in direct search experiments imposes a very strong constraint on DM models, ruling out or severely constraining most of the 
simplest single-component frameworks. It is therefore very important to study the 
constraints imposed on the pNBG parameter space by direct search data. Direct search reaction for the pNBG DM is mediated by Higgs boson in $t-$ channel 
as in Higgs portal scalar singlet DM. Spin-independent direct search cross-section for pNBG DM is given by:
 \begin{equation}
   \sigma_{pNBG}^{SI}=\frac{\alpha_n^2 \mu_{n}^2}{4\pi m_{G_S}^2},\quad \mu_n=\frac{m_n m_{G_S}}{m_n+m_{G_S}},
    \end{equation}
 where
  \begin{eqnarray}
  \alpha_n &=& m_n\sum_{u,d,s} f_{T_q}^{(n)} \frac{\alpha_q}{m_q} + \frac{2}{27} f_{T_g}^{(n)} \sum_{q=c,t,b}\frac{\alpha_q}{m_q}, \nonumber \\
  &=& \frac{m_n \lambda^{''}}{m_h^2}[(f_{T_u}^{(n)}+f_{T_d}^{(n)}+f_{T_s}^{(n)})+\frac{2}{9}(f_{T_u}^{(n)}+f_{T_d}^{(n)}+f_{T_s}^{(n)})].
  \end{eqnarray}
  The subindex $n$ refers to the nucleon (proton or neutron). We use default form factors for proton for calculating direct search cross-section: 
 $f_{T_u}^p=0.0153$ , $f_{T_d}^p=0.0191$ , $f_{T_s}^p=0.0447$\cite{Alarcon:2011zs,Alarcon:2012nr}. In Fig. \ref{fig:DD1}, 
we show the spin-independent nucleon-pNBG DM cross-section for the chosen benchmark point, plotted as a function of DM mass 
$m_{G_S}$ . The green points in this figure also meet the  relic density constraint. Bounds from LUX 2016\cite{Akerib:2016vxi}, Xenon1T\cite{Aprile:2017iyp} and PANDA X\cite{Cui:2017nnn} and future predictions from XENONnT\cite{Aprile:2015uzo} are also in the figure. Clearly, the figure shows that the degenerate case is excluded by the direct search bound except in the Higgs resonance region. This is simply because for 15 degenerate DM particles, the values of $\lambda^{''}$ required 
to satisfy relic density constraint correspond to a direct detection cross section large enough to be excluded by the data (except in the 
resonance region). 

\subsubsection{Non-degenerate pNBGs}

We next consider the model when the pNGBs are not degenerate; we will see that in this case the allowed parameter space is considerably enlarged. 
For this it is sufficient to consider cases where there is a single lightest pNGB, and the simplest situaiont in which this occurs is when 
$m_1=m_2=m_3=m$ and $m_4\neq m$. In this case the pNGB spectrum is
\begin{equation}
\begin{array}{c|c|l}
\mbox{type} & \mbox{\#~of~degenerate~pNGBs} & \mbox{mass}  \cr \hline
A & 8 & m_A=2m \cr
B & 6 & m_B=m + m_4 \cr
C & 1 & m_C=(3 m_4+m)/2
\end{array}
\end{equation}
and there are two cases:
\begin{align}
I:&~~ m>m_4: \quad\Rightarrow \quad  m_A > m_B > m_C \cr
II:&~~m<m_4: \quad\Rightarrow \quad  m_A < m_B < m_C 
\end{align}
In the first case there is a single lightest pNGB (type $C$), while in the second there are 8 lightest states (type $A$). 

\begin{figure}[h]
 \includegraphics[height=5.5cm]{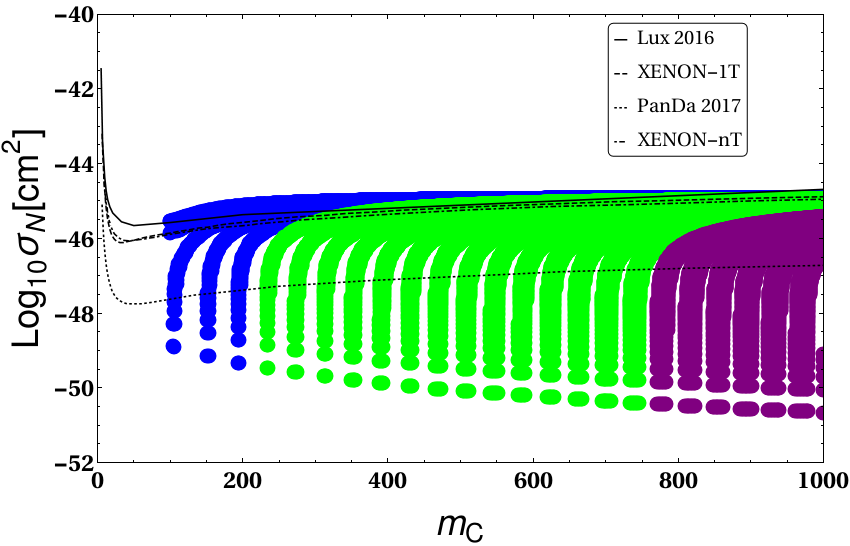}
 \includegraphics[height=5.5cm]{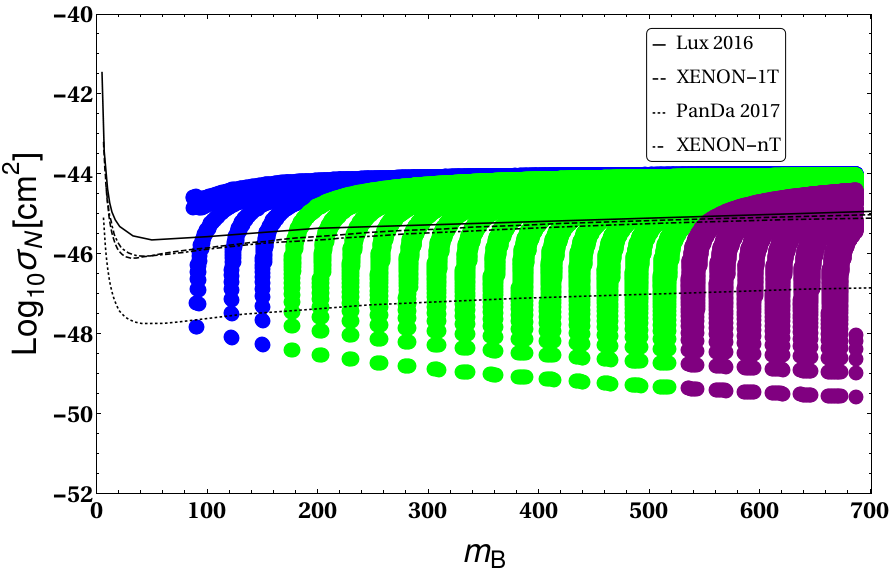}
  \caption{Spin independent direct detection cross section for $C$ (left) and $B$ (right) for relic density satisfied region have been 
compared with LUX 2016\cite{Akerib:2016vxi} and XENON 1T\cite{Aprile:2017iyp}, Panda X\cite{Cui:2017nnn} experimental constraints and with expected sensitivity from XENONnT\cite{Aprile:2015uzo}. Different choices of $\kappa_1$ are shown in different colours: $0.01-0.25$ (blue), $0.25-0.45$ (green), $0.45-1.0$ (purple)}
  \label{fig:DD-nondeg}
  \end{figure}

Note that due to the presence of three types of pNGBs, the total relic density should be written as 
$\Omega_T = n_A \Omega_A + n_B \Omega_B + n_C \Omega_C$, where $n_{A/B/C}$ are the number of 
degeneracies of the respective species.  Compared to the case with all degenerate pNGBs, here the coupling 
$\kappa_1$ also comes into play along with $\lambda$. In case of all degenerate pNGBs, we have seen that 
the relic density satisfied region was ruled out by the direct detection cross-section limits excepting for Higgs resonance. This is because a 
large $\lambda$, as required to satisfy relic density, makes the direct detection cross section significantly higher 
than the experimental limits. Here we can make $\lambda$ relatively free as $\kappa_1$ also 
enters in the game. which does not affect the direct detection cross section. We can allow a further 
smaller value of $\lambda$ in this non-degenerate case, provided not all 15 pNGBs may not effectively contribute to 
the relic density. This can 
happen once we put mass of one type of pNGBs (out of A,B and C) near the resonance region 
$\sim m_h/2$ where the annihilation cross-section is large to make the corresponding relic density very small. In 
this case, the total relic density gets contribution from the two remaining types of pNGBs, therby a smaller 
$\lambda$ (compared to all degenerate case) can be chosen. Note that case II would be more promising compared to 
case I from this point of view as by putting $m_A \simeq m_h/2$, we effectively have remaining 7 pNGBs to contribute 
to relic. In Fig.\ref{fig:DD-nondeg}, we demonstrate
the masses, number of degeneracies and possible interactions among the pNGB DM candidates for the this case II.  
In this case $ m_B \simeq ( 4 m_C + m_h)/6$ and $ m_C > m_h/2 $ and the relic density reads (since $ \Omega_A \simeq 0 $), 
\begin{align}
\Omega_T\simeq\Omega_C+6\Omega_B,
\end{align} 
with $0.1175<\Omega_T<0.1219$ following PLANCK data\cite{Ade:2015xua}.
\begin{figure}[thb!]
$$
\includegraphics[height=6.5cm]{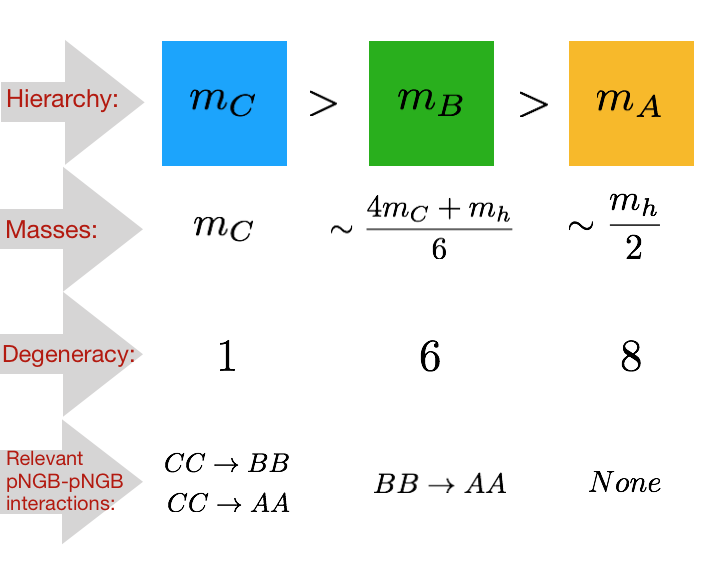}
$$
\caption{Masses, degeneracies and possible interactions of pNGB DMs in phenomenologically viable non-degenrate case, illustrated in this analysis.}
\label{fig:Dashen_chart}
\end{figure}

In order to obtain $ \Omega_{C,B}$ we consider the set of coupled Boltzmann equations for the $C,\,B_i$ and $A_i$ number densities (the last included for completeness):
\begin{align}
\frac{dn_C}{dt}+3Hn_C&=-\langle\sigma v\rangle_{G_C G_C\rightarrow{\textrm{SM}}}\left(n_C^2-n_C^{\textrm{eq}2}\right)-
6\langle\sigma v\rangle_{G_C G_C\rightarrow G_BG_B}\left(n_C^2-\frac{n_C^{\textrm{eq2}}}{n_B^{\textrm{eq}2}}n_B^2\right)\nonumber\\
&-8\langle\sigma v\rangle_{G_C G_C\rightarrow G_AG_A}\left(n_C^2-\frac{n_C^{\textrm{eq2}}}{n_A^{\textrm{eq}2}}n_A^2\right), \nonumber \\
\frac{dn_{B_i}}{dt}+3Hn_{B_i}&=-\langle\sigma v\rangle_{G_{B_i} G_{B_i}\rightarrow{\textrm{SM}}}\left(n_{B_i}^2-n_{B_i}^{\textrm{eq}2}\right)-
8\langle\sigma v\rangle_{G_{B_i} G_{B_i}\rightarrow G_{A_i}G_{A_i}}\left(n_{B_i}^2-\frac{n_{B_i}^{\textrm{eq2}}}{n_{A_i}^{\textrm{eq}2}}n_{A_i}^2\right) \nonumber\\
&+\langle\sigma v\rangle_{G_C G_C\rightarrow G_{B_i} G_{B_i}}\left(n_C^2-\frac{n_C^{\textrm{eq2}}}{n_{B_i}^{\textrm{eq}2}}n_{B_i}^2\right), \nonumber \\
\frac{dn_{A_i}}{dt}+3Hn_{A_i}&=-\langle\sigma v\rangle_{G_{A_i} G_{A_i}\rightarrow{\textrm{SM}}}\left(n_{A_i}^2-n_{A_i}^{\textrm{eq}2}\right)+
\langle\sigma v\rangle_{G_C G_C\rightarrow G_{A_i}G_{A_i}}\left(n_C^2-\frac{n_C^{\textrm{eq2}}}{n_{A_i}^{\textrm{eq}2}}n_{A_i}^2\right)+\nonumber\\
&\langle\sigma v\rangle_{G_B G_B\rightarrow G_{A_i}G_{A_i}}\left(n_B^2-\frac{n_B^{\textrm{eq2}}}{n_{A_i}^{\textrm{eq}2}}n_{A_i}^2\right).
\label{eq:BEQ-nondeg}
\end{align}
In above we have ignored the interactions with the LSP.  The numerical factors correspond to the number of final state particles  each pNBG species can annihilate to. As is evident, a crucial role is played by the DM-DM contact interactions generated by (see Eq. \ref{scalar})
\begin{align}
V_{\textrm{pNBG}}^{\textrm{Int}}=\frac{\kappa_1^2}{4}\sum_{a,b} (G_S^a)^2(G_S^b)^2.
\end{align}

Since we assume that the $A$ mass lies in the Higgs resonance region, $\langle\sigma v\rangle_
{G_A G_A\rightarrow{\textrm{SM}}}$ is very large, and produces very small relic density which we neglect in the following estimates; one can easily calculate that with $m_A\sim m_h/2$, the $A$ contributes less than 1\%\ to the total 
relic if $\lambda^{''}>10^{-3}$. We will use this value of $\lambda^{''}$ 
as a lower limit in our analysis.  

Even if we neglect contributions from A type, pNBG, Eqs.(\ref{eq:BEQ-nondeg}) are coupled and must be solved
 numerically to find the freeze-out
of the individual components. However, as it is shown in \cite{Bhattacharya:2016ysw}, for interacting multicomponent DM
scenario\cite{Bhattacharya:2013hva,Ahmed:2017dbb}, annihilation of heavier components to lighter ones are crucial in determining 
the relics of only heavier components, while for lighter components it has mild effect. Hence we can derive an approximate 
analytic expressions for the individual relic densities by considering the annihilation of one pNGB kind to the lighter species. In this case we find
\begin{align}
\Omega_Ch^2&\sim \frac{2.0\times 10^{-10}~{\rm GeV^{-2}}}{\langle\sigma v\rangle_{G_C G_C\rightarrow{\textrm{SM}}}
+6\langle\sigma v\rangle_{G_C G_C\rightarrow G_BG_B}+
8\langle\sigma v\rangle_{G_C G_C\rightarrow G_AG_A}}, \nonumber \\
\Omega_Bh^2&\sim\frac{2.0 \times 10^{-10}~{\rm GeV^{-2}}}{\langle\sigma v\rangle_{G_B G_B\rightarrow{\textrm{SM}}}+8\langle\sigma 
v\rangle_{G_B G_B\rightarrow G_AG_A}}, \nonumber\\
\Omega_T&=\Omega_C+6\Omega_B.
\label{relic:BC}
\end{align}
These approximate analytical results are in reasonably good agreement with the numerical solutions,
 as we show in Appendix III.
 
\begin{figure}[htb!]
$$
\includegraphics[height=5.5cm]{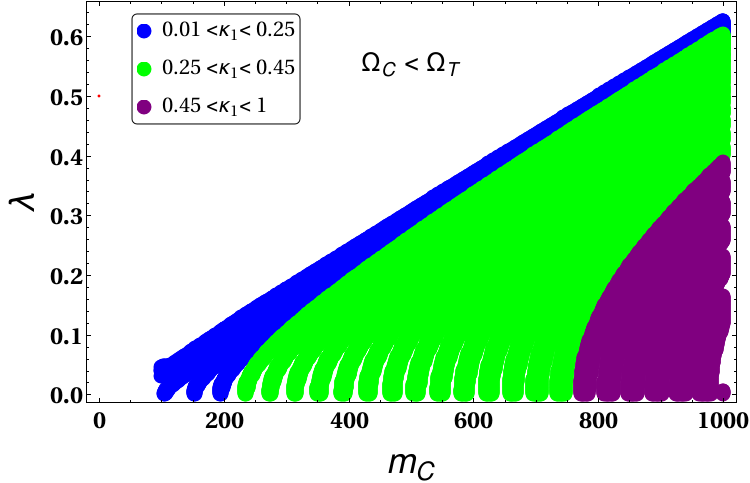}
\includegraphics[height=5.5cm]{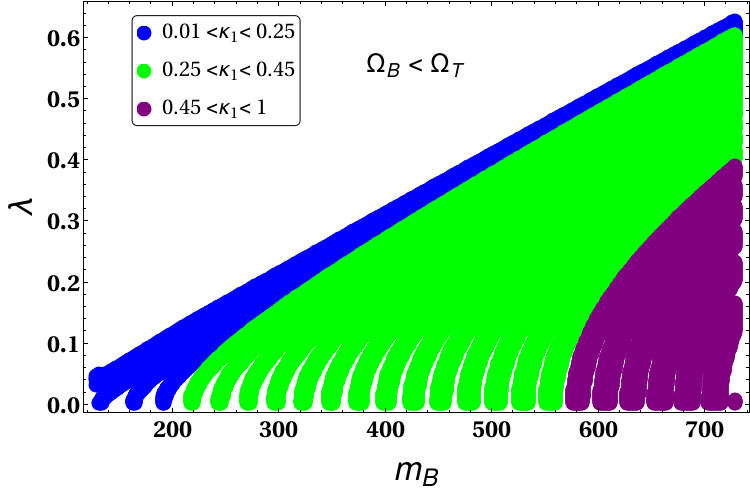}
$$
$$
\includegraphics[height=5.5cm]{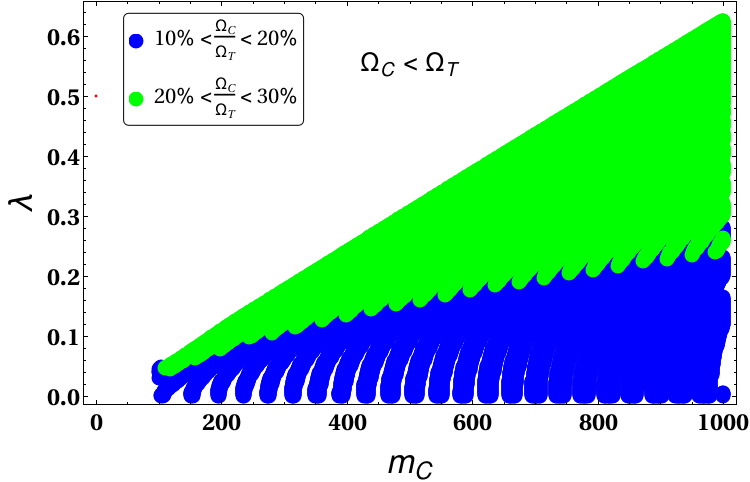}
\includegraphics[height=5.5cm]{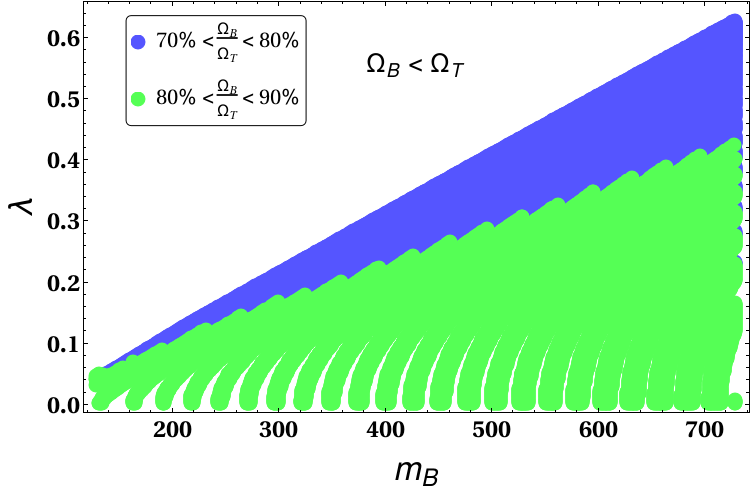}
$$
  \caption{ $\lambda$ vs $m_C$ (left) and $\lambda$ vs $m_{B}$ (right) for satisfying relic density $0.1175<\Omega_T<0.1219$.[Top panel:] Different choices
  of $\kappa_1$ are shown in different colours: $0.01-0.25$ (blue), $0.25-0.45$ (green), $0.45-1.0$ (purple). [Bottom panel:]
Relative contributions of individual DM candidates to totla relic abundance have been shown in different colors. }
  \label{fig:beta1}
  \end{figure}

 The DM phenomenology here crucially depends on the couplings $ \lambda$ and $\kappa_1$, which we have 
varied freely for the scan. In Fig.~\ref{fig:beta1} [top panel], we show the allowed parameter space in the $\lambda-m_C$ (left)
 and $\lambda-m_B$ (right) planes for $\kappa_1$ varying between $0.01-0.25$ (blue), $0.25-0.45$ (green), $0.45-1$ (purple),  that satisfies 
individually $\Omega_C<\Omega_T$ (left) and $\Omega_B<\Omega_T$ (right). It is observed that  larger values of $\kappa_1$ requires also larger 
DM masses to produce the required annihilation cross-section. In Fig.~\ref{fig:beta1} [bottom panel] 
we show the relative contributions to total relic density by individual components
 in $\lambda-m_C$ (left) plane and $\lambda-m_B$ (right) plane for varying $\kappa_1$ from 0.01 to 1.

\begin{figure}[htb!]
$$
 \includegraphics[height=5.4cm]{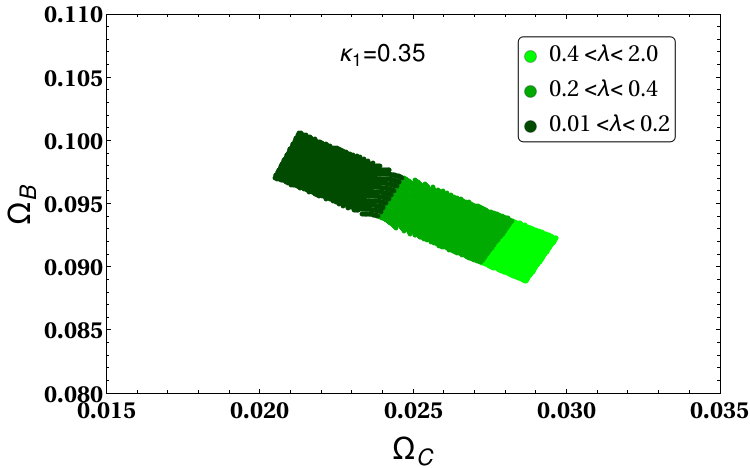}
 \includegraphics[height=5.4cm]{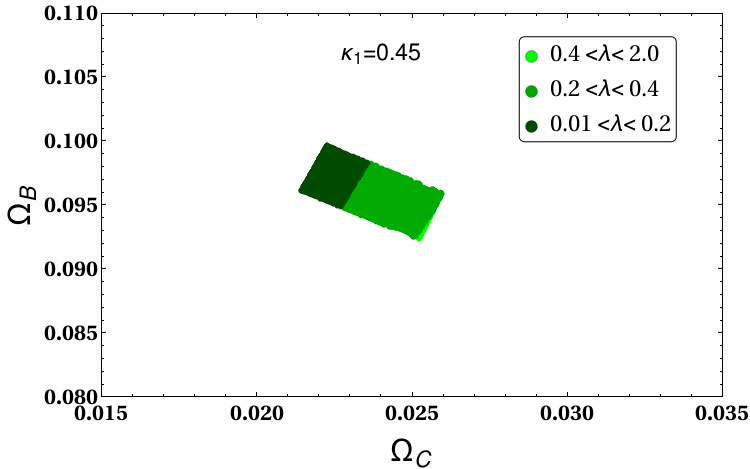}
$$
$$
 \includegraphics[height=5.4cm]{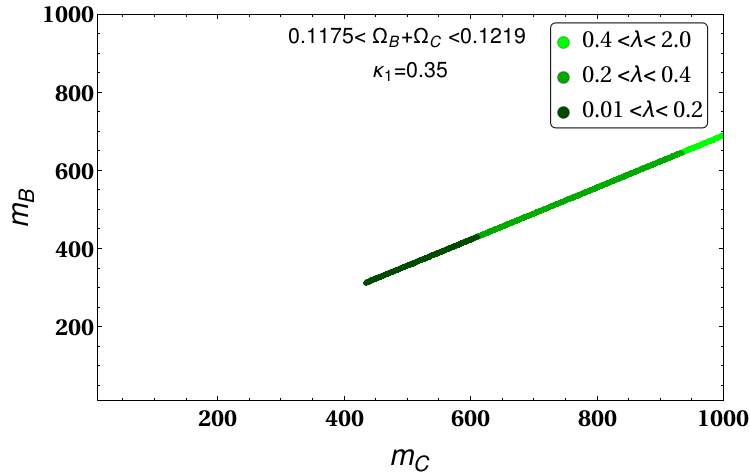}
 \includegraphics[height=5.4cm]{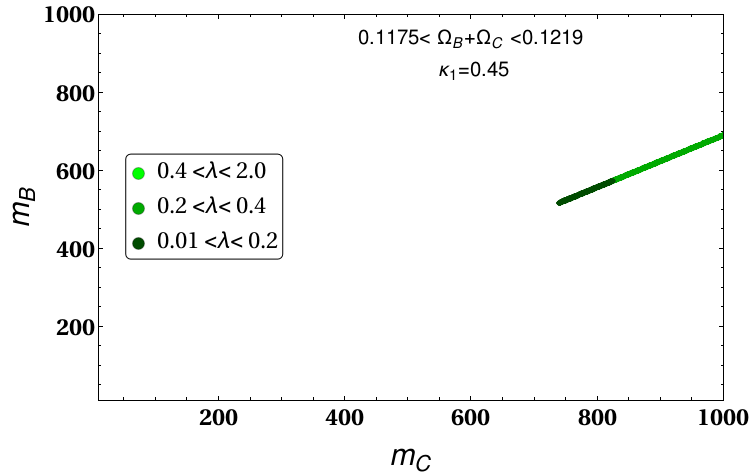}
$$
  \caption{[Top panel:] Contribution to $\Omega_T$ from $\Omega_C$ and $\Omega_B$ for
different choices of $\kappa_1=0.35$ (left) and 0.45 (right). Different ranges of $\lambda$ are indicated $\lambda=\{0.01-0.2\}$ (dark green), $\{0.2-0.4\}$ (green)
 and $\{0.4-2.0\}$ (lighter green) respectively. [Bottom panel:] Mass correlation $(m_B-m_C)$  in allowed relic density parameter space. Color codes remain the same
 as in top panel.}
  \label{fig:beta2}
  \end{figure}

In Fig.~\ref{fig:beta2}, we also show the relative contribution of relic density of one type of DM for a fixed $\kappa_1$; on the left (right) side, we choose $\kappa_1=0.35$  ($\kappa_1=0.45$). Contributions from different values of
$\lambda$ are shown in different colours. We note that
 $\Omega_C$ yields the dominant contribution to $\Omega_T$, which occurs because $C$ annihilation cross section is larger than for the $B$, by the contribution from the $CC\to BB$ process and also due to the larger degeneracy (6) of B component. We also see that with larger $\kappa_1$, the regions of larger $\lambda$ disappear ({\it i.e.} are inconsistent with the constraints).

%LUX\cite{Akerib:2016vxi}, Xenon1T\cite{Aprile:2017iyp} and PANDA X\cite{Cui:2017nnn} are shown with the expected sensitivity from XENONnT\cite{Aprile:2015uzo}
Finally we show spin independent direct search cross section of C and B-types of DM in Fig \ref{fig:DD-nondeg}. A large region of parameter space is allowed by the LUX limit; this is because the DM-DM conversion allows different pNBG species meet the required relic density, without contributing to direct search cross sections. 
Clearly depending on how large one can choose $\kappa_1$, the mass
of the DM gets heavier to satisfy relic density and direct search constraints. Also due to the larger DM-DM conversion cross-section, the direct-detection probability for the $C$ type of pNBG is smaller than for the $B$ type.

We conclude the section by discussing a specific region of parameter space of the MSSM where the LSP is dominated by the wino/wino-Higgsino component, and also contributes negligibly to relic density. The LSP has four different contributions from the two Higgsinos ($\tilde H_{u,d}$), wino ($\tilde W$) and bino ($\tilde B$):\cite{Haber:1984rc,Drees:2004jm,Wess:1992cp,Martin:1997ns}
\begin{align}
  \chi_0=Z_{11}\tilde{B}+Z_{12}\tilde{W}+Z_{13}\tilde{H_u}+Z_{14}\tilde{H_d}
\end{align}
  where the $Z_{1j}$ represent mixing angles.
Now, since we are interested in the case where the DM density is dominated by the pNBGs, the relic density of neutralino ($\Omega_{\chi_0}$) has to be very small.
This is possible when the neutralino is generally dominated by the wino component or wino Higgsino components\cite{Profumo:2017ntc}.
 We tabulize two such examples (Table \ref{Tab:RL}) where we assume squarks, sleptons and gluinos of the order of 2 TeV, $\tan \beta=5$ and all 
trilinear couplings at zero, excepting $A_t=-1000$ to yield correct Higgs mass. We find that the contribution of the LSP (neutralino)
 to the relic density is small and also the spin independent direct detection cross section $\sigma^{\textrm{SI}}_{\chi_0\chi_0\rightarrow N N}$
 is well below the PANDA X\cite{Cui:2017nnn} experimental limit.

 \begin{table}
\begin{center}
\begin{tabular}{ |c|c|c|c|c|c|c|c|c|c| } 
 \hline
 $\mu$ & $M_1$ & $M_2$  & $Z_{11}$& $ Z_{12}$ & $Z_{13}$ & $Z_{14}$ & $m_{\chi_0}$ & $\Omega_{\chi_0}h^2$ & $\sigma^{\textrm{SI}}_{\chi_0\chi_0\rightarrow N N}$\\ 
\hline
 250  & 3000  & 200  & 0.009 & 0.672 & 0.568 & 0.475 & 250  & $2\times 10^{-4}$ & $1.2\times 10^{-46}$ \\ 
\hline
 700 & 3000 & 400 & 0.003 & 0.976 & 0.178 & 0.124 & 400 & $5.6\times 10^{-3}$ & $5.4\times 10^{-46}$  \\ 
 \hline
\end{tabular}
\end{center}
\caption{Relic density and corresponding SI direct detection cross section for wino/wino-Higgsino dominated neutralino with 
$\tan\beta=5$. The input parameters in terms Bino ($M_1$), Wino ($M_2$) and Higgsino ($\mu$) masses and the output in terms of neutralino mass and mixing parameters are indicated. All the masses are in GeVs. Spin independent direct search cross section is in cm$^2$.}
\label{Tab:RL}
\end{table}

 %for example, if $\mu=400, M_1=500, M_2\in\{300,\,500\}, M_3=800$ (all in GeV), with squarks, sleptons and gluinos of the order of 1 TeV,
%$\tan \beta=2$ and all trilinear couplings at zero, excepting $A_t=-1000$ to yield correct Higgs mass. Within this region we find $Z_{12} \in\{0.884,\,0.552\}$ 
%while $Z_{13}\simeq Z_{14} \in  \{0.3,\,0.6\}$, while the contribution of the LSP to the relic density is small: $\Omega_{\rm LSP} \in \{0.003,\,0.01\}$, corresponding
%to only 3\% to 10\% of the total. Note that while the wino generally dominates, the higgsino component may not always be ignored.

 %%%%%%%%%%%%%%%%%%%%%%%%%%%%%%%%%%%%%% 
\subsection{Non-negligible pNBG-LSP interaction limit}
%%%%%%%%%%%%%%%%%%%%%%%%%%%%%%%%%%%%%%%

In this section we will consider some of the effects of the pNGB-LSP couplings; for simplicity we will assume that the fifteen pNGBs are degenerate 
(it is straightforward to relax this assumption).  As noted above, there are cases where the LSP receives a non-negligible contribution from the 
Higgsinos, in which case the LSP-pNGB interactions cannot be ignored, even though the DM relic density is still dominated by the pNBGs. 
In this case the evolution of the pNBG density is described by
\begin{align}
\dot{n_{G_S}}+3Hn_{G_S}&=-\langle \sigma v \rangle_{G_SG_S\to SM} (n_{G_S}^2-n_\chi^{\rm eq}{}^2)- \langle \sigma v \rangle_{G_SG_S \to \chi \chi} (n_{G_S}^2-n_\chi^{\rm eq}{}^2), \nonumber \\
&= -\left[ \langle \sigma v \rangle_{G_SG_S\to SM}+\langle \sigma v \rangle_{G_SG_S \to \chi \chi} \right](n_{G_S}^2-n_\chi^{\rm eq}{}^2).
\label{eq:case-b}
\end{align}
where in the last term, we used $n_{\chi_0}=n_{\chi_0}^{\rm eq}$ for the neutralinos since in the parameter region being considered they interact sufficiently 
strongly with the standard model to ensure they are in equilibrium; the decoupling of the LSP from the SM occurs much later than the decoupling of the pNBGs. 
We also assumed here that pNGBs are heavier than MSSM neutralino.
Using then standard techniques\cite{McDonald:1993ex} we find that the DM relic abundance is given approximately by 
 \begin{align}
 \Omega_T h^2= 15\times\frac{2.0\times 10^{-10} \textrm{GeV}^{-2}}{\langle \sigma v\rangle_{G_SG_S \to SM}+\langle \sigma v\rangle_{G_SG_S \to \chi_0 \chi_0}},
 \end{align}
 where the presence of the second term in the dominator will be instrumental in accommodating the direct detection constraints and the numerical factor 15
to take care of fifteen degenerate pNGB species. 
 
\begin{figure}[htb!]
$$
 \includegraphics[height=5.5cm]{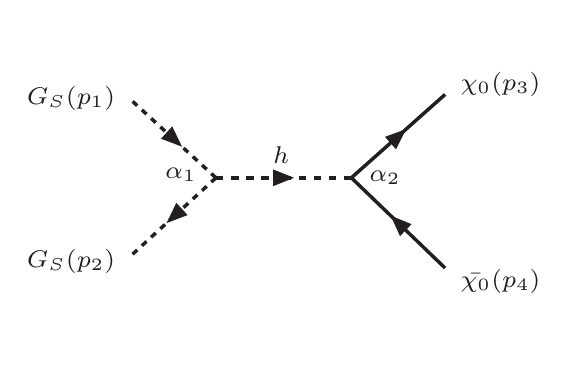}
$$
\caption{Feynman graph for Neutralino-pNBG interaction}
\label{fig:fdcon}
\end{figure}

The Feynman graph responsible for the pNBG-LSP interactions is presented in Fig.~\ref{fig:fdcon}. The two vertex factors are named as $\alpha_1=2\lambda^{\prime\prime}v_d$ and $\alpha_2$ (elaborated below). This leads to the following annihilation
cross section evaluated at threshold $s=4m_{G_s}^2$: 
\begin{align}
(\sigma v)_{G_sG_s \to \chi \chi}|_{s=4m_{G_s}^2}=&\frac{|\lambda''\textrm{ } v_d\textrm{ } \alpha_2\textrm{ } C|^2}{8\pi}\frac{2-(m_{\chi_0}/m_{G_s})^2}
{ (4m_{G_s}^2-m_h^2)^2} \left\{ 2-\frac{m_{\chi_0}^2}{m_{G_s}^2} \left[1 + \textrm{Re}\left(\frac{C^2}{|C|^2} \right)\right] \right\},
\end{align}
where $m_{\chi_0}$ is the neutralino mass and 
\begin{align}
&\lambda^{\prime\prime}v_d=\frac{\lambda v_d}{2\cos\beta}, \nonumber \\
&C = \left(Z_{14}\sin\beta - Z_{13}\cos\beta \right)\left(Z_{12}-\tan\theta_W Z_{11} \right)\label{Eq:C}.
\end{align}
The vertex containing $\alpha_2$ is generated by the Higgs-neutralino interaction\cite{Martin:1997ns}:
\begin{align} 
-i g_2\bar{\chi}_0(C^* P_L+C P_R)\chi_{0} h, 
\end{align} 
where $g_2$ is the $SU(2)_L$ gauge coupling constant; as previously, we assumed $ \alpha = \beta + \pi/2 $. The detailed calculation of cross-section is
illustrated in Appendix II.

\begin{figure}[htb!]
$$
 \includegraphics[height=5.5cm]{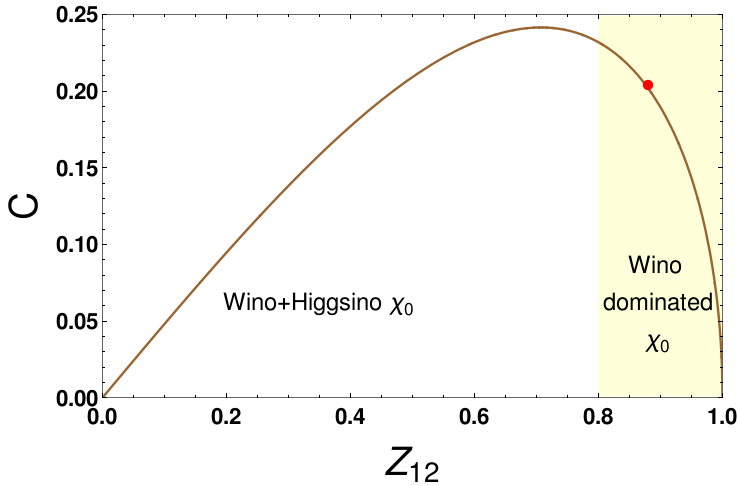}
$$
\caption{A plot between $C$ and $Z_{12}$ for $\tan\beta=5$ assuming wino or wino-Higgsino dominated neutralino with $Z_{11}=0$ and $Z_{13}=Z_{14}$. Our choice ($Z_{12}=0.88$ and $C=0.2$) has been denoted by a red dot. }
\label{fig:coup}
\end{figure}

We perform a scan of the pNBG DM parameter space by varying DM mass and coupling of pNGB with SM (proportional to $\lambda$) with pNBG DM-Neutralino annihilations into
account. For that we choose two benchmark points by fixing $\textrm{tan}\beta=5$ and using $\alpha=\beta+\frac{\pi}{2}$ as mentioned in Table \ref{Tab:BP2}.
  This is following from what we obtained from Table \ref{Tab:RL}, where neutralino has minimal relic density, but sizeable pNGB-neutralino interaction.
 The two interaction coefficients ($\lambda^\prime$ and $\lambda^{\prime \prime}$) in this approximation turns out to be:
\begin{align}
\lambda^{\prime}=& \frac{\lambda}{2} ~\textrm{sin}\alpha~\textrm{cos}\alpha=\frac{0.195}{2}~\lambda, \nonumber\\
\lambda^{\prime\prime}=& \frac{\lambda}{2}~ v_d ~\textrm{cos}\alpha~(\textrm{tan}\alpha-\textrm{tan}\beta)=-\frac{247}{2}~\lambda.
\end{align}

\begin{table}[h]
\hspace{-8mm}
\begin{center}
\begin{tabular}{ |c|c|c|c|c|c|c| } 
 \hline
 $M_A$ (GeV) & $M_{\chi_0}$ (GeV) & $C$ & $Z_{11}$ & $Z_{12} $ & $Z_{13}$ & $Z_{14}$ \\ 
\hline
 400  & 200  & 0.2 & 0 & 0.33 & 0.33 & 0.88 \\ 
\hline
 600 & 400  & 0.2 & 0 & 0.33 & 0.33 & 0.88  \\ 
 \hline
\end{tabular}
\end{center}
\caption{Benchmark points used to find out the relic density and SI direct detection cross section of pNGB DM using pNGB-LSP interaction.
 We mention physical masses of neutralino ($\chi_0$) and pseudoscalar Higgs $A$ and mixing pattern of neutralinos.}
\label{Tab:BP2}
\end{table}

%\begin{align}
%M_A=& 400~(600) ~\textrm{GeV},\textrm{   } \textrm{tan}\beta=5,\textrm{   }\alpha=\beta+\frac{\pi}{2} \nonumber \\
%\lambda^{\prime}=& \lambda \textrm{sin}\alpha\textrm{cos}\alpha=0.195\lambda \nonumber \\
%\lambda^{\prime\prime}=& \lambda v_d \textrm{cos}\alpha(\textrm{tan}\alpha-\textrm{tan}\beta)=-247\lambda \nonumber \\
%M_{\chi}=& 200~ (400)~\textrm{GeV}, ~~C_{L1}=C_{R1}=0.2 
%\end{align} 

Now as we are working with the wino or wino-Higgsino dominated neutralino, it is safe to consider $Z_{11}=0$. For simplification purpose we also assume
$Z_{13}=Z_{14}$. Next we employ the constraint $|Z_{11}|^2+|Z_{12}|^2+|Z_{13}|^2+|Z_{14}|^2=1$. In that case $C$ in Eq.(\ref{Eq:C}) turns out to be a function of only $Z_{12}$. In Fig.~\ref{fig:coup} a line plot has been presented to show the variation of $C$ with $Z_{12}$. For our analysis, we choose $C=0.2$ and $Z_{12}=0.88$ (wino dominated LSP) 
which has been denoted by a red dot in Fig.~\ref{fig:coup}.

\begin{figure}[htb!]
$$
\includegraphics[height=5.2cm]{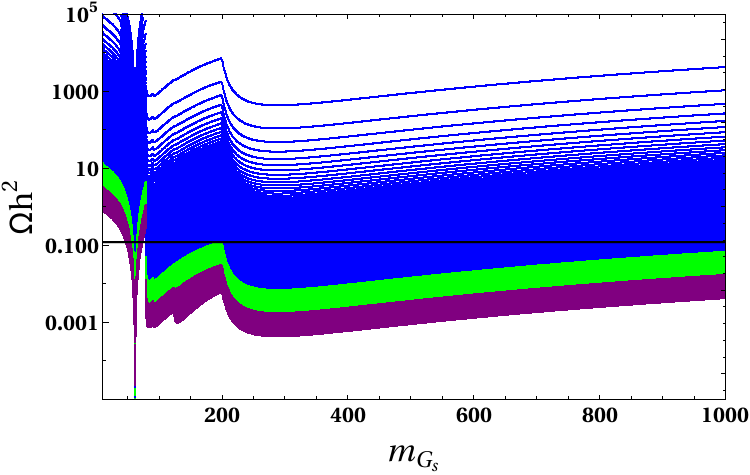}
 \includegraphics[height=5.2cm]{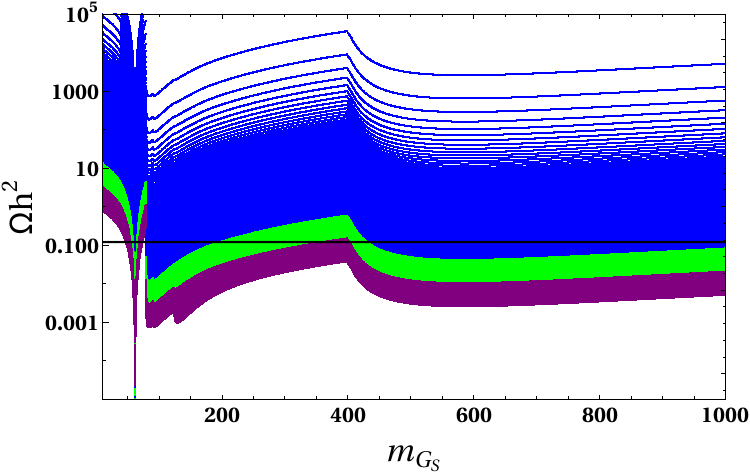}
$$
\caption{Relic density as a function of $m_{G_S}$ for $\lambda=0.005-0.25$ (blue), $0.25-0.5$ (green) and $0.5-2$ (purple). 
We assumed all 15 pNBGs are degenerate and included the effects of the pNBG-LSP interaction with $\tan\beta=5$,  $C=0.2$, and a 
neutralino mass of $m_{\chi}=200$ GeV (left) or $m_{\chi}=400$ GeV (right).}
\label{fig:omega-MSSM}
\end{figure}

\begin{figure}[htb!]
$$
\includegraphics[height=5.5cm]{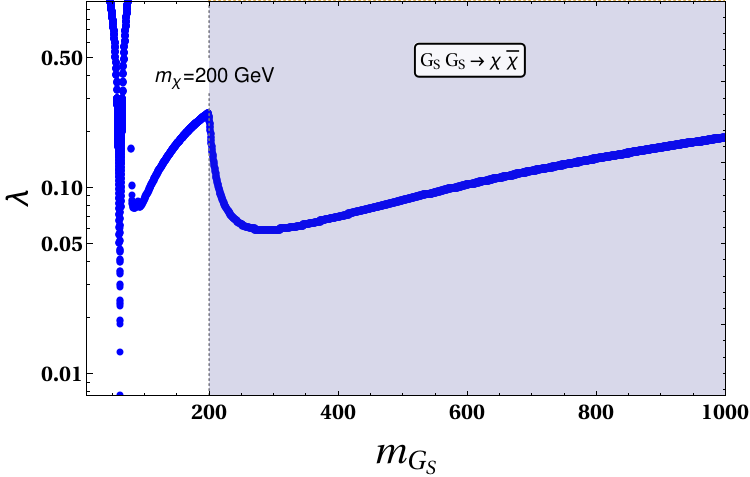}
 \includegraphics[height=5.5cm]{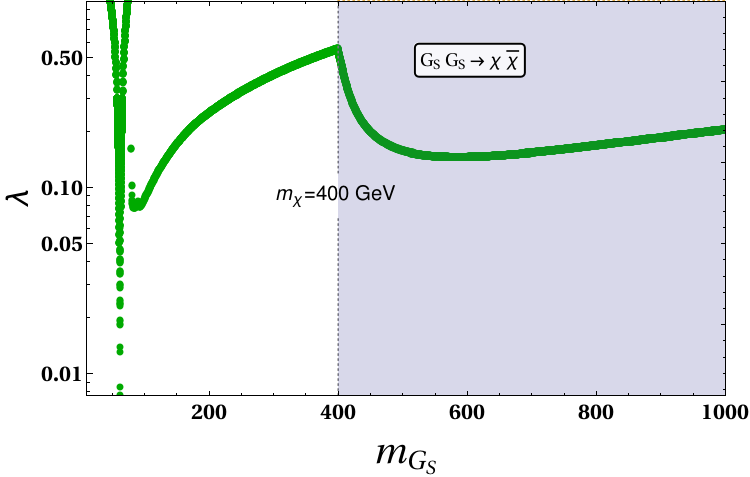}
$$
\caption{Region in the $ m_{G_S} - \lambda $ space allowed by the relic density constraint. We assumed all 15 pNBGs are degenerate
and included the effects of the pNBG-LSP interaction with $\tan\beta=5$,  $C=0.2$, and a neutralino mass of $m_{\chi}=200$ GeV (left) or $m_{\chi}=400$ GeV (right).}
\label{fig:M-K-MSSM15}
\end{figure}

%\begin{figure}[htb!]
%$$
%% \includegraphics[height=5.5cm]{Omega_1_200.png}
%% \includegraphics[height=5.5cm]{Omega1_400.png}
%$$
%\caption{Relic density as a function of \jw{$m_{G_S}$ for $\lambda=0.005-0.25$ (blue), $0.25-0.5$ (green) and $0.5-2$ (purple). We assumed a single lightest pNBG  and included the effects of the pNBG-LSP interaction with $\tan\beta=5$,  $C=0.2$, and a neutralino mass of $m_{\chi}=200$ GeV (left) or $m_{\chi}=400$ GeV (right)}}
%\label{fig:Omega-MSSM}
%\end{figure}
%
%\begin{figure}[htb!]
%$$
%% \includegraphics[height=5.5cm]{mass_coupl-200-1.png}
%% \includegraphics[height=5.5cm]{mass_coup-400-1.png}
%$$
%\caption{\jw{Region in the $ m_{G_S} - \lambda $ space allowed by the relic density constraint. We assumed one lightest pNBG and included the effects of the pNBG-LSP interaction with $\tan\beta=5$,  $C=0.2$, and a neutralino mass of $m_{\chi}=200$ GeV (left) or $m_{\chi}=400$ GeV (right)}}
%\label{fig:M-K-MSSM}
%\end{figure}

\begin{figure}[htb!]
$$
 \includegraphics[height=5.5cm]{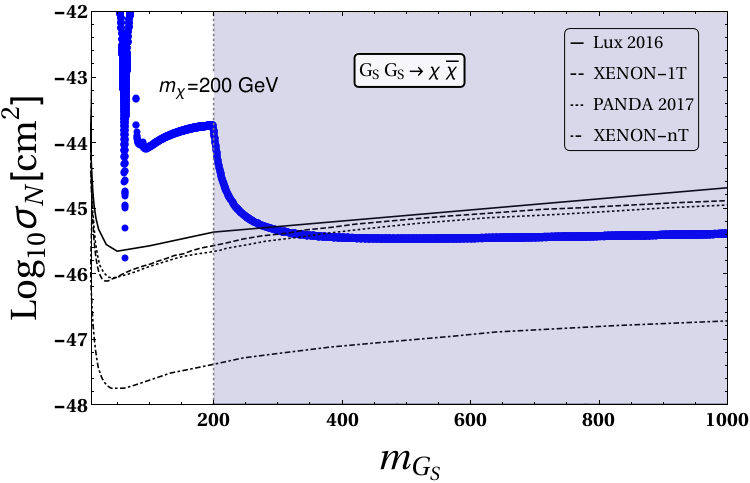}
 \includegraphics[height=5.5cm]{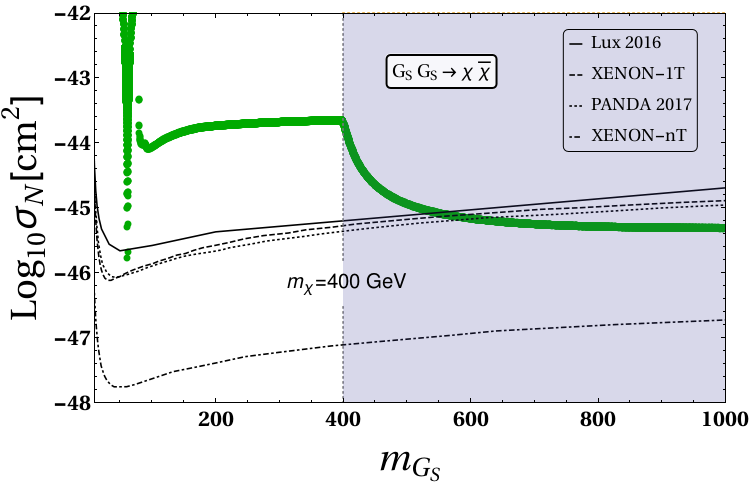}
$$
\caption{Direct search constraints for pNBG DMs from LUX 2016\cite{Akerib:2016vxi}, XENON 1T\cite{Aprile:2017iyp}, PANDA X\cite{Cui:2017nnn} and future prediction
of XENONnT\cite{Aprile:2015uzo} for the case of 15 degenerate
pNBGs where the effects of the pNBG-LSP interaction is considered with $\tan\beta=5$, $C=0.2$ and neutralino mass $m_{\chi}=200$ GeV (left), $m_{\chi}=400$ GeV (right). }
\label{fig:direct-MSSM}
\end{figure}

The relic density for the case of 15 degenerate pNBGs as a function of pNBG DM mass is shown in Fig.~\ref{fig:omega-MSSM}, which takes
into account the effects of the pNBG-LSP interaction. This is scanned for $\lambda=0.005-0.25$ (blue), $0.25-0.5$ (green) and $0.5-2$ (purple).
The graph clearly shows flattening of relic density lines (which corresponds to
fixed values of $ \lambda $)  %\jwn{Do these lines correspond to fixed values of $ \lambda$?} 
for $m_{G_S} \ge m_{\chi}$, when the pNBG$\to $LSP
annihilation channel becomes kinematically allowed. The allowed parameter space for the pNBGs in mass-coupling plane is shown in Fig.~\ref{fig:M-K-MSSM15}. 

In addition to predicting the expected relic density, the model should also comply with the constraints from direct detection experiments.
The restrictions imposed by the LUX 2017\cite{Akerib:2016vxi}, XENON1T\cite{Aprile:2017iyp} and PANDA X\cite{Cui:2017nnn} experiments are presented in Fig.~\ref{fig:direct-MSSM} for the cases
of 15 degenerate pNBGs with sizeable interaction with neutralinos.  The results clearly indicate that  in the presence of pNBG-LSP interactions the direct detection impose milder restrictions on parameter space compared to those cases where this coupling is negligible (compare figures \ref{fig:DD1} and \ref{fig:direct-MSSM}). In particular, the fully degenerate pNGB case can now comply with the direct search constraints. 
It is again worth emphasizing the role played by the neutralino: it provides a new channel through which the 
pNBGs can annihilate (pNBG$\to$LSP) that does not affect the direct detection cross section, this allows smaller values of $ \lambda $ (thus relaxing direct detection constraints), while keeping 
a large enough annihilation cross section, needed to meet the relic abundance requirements. As noted earlier, this occurs in the region where the LSP is wino/wino Higgsino dominated and in this reigon of parameter space $ \Omega_{\rm LSP} \ll \Omega_{\rm pNBG} \simeq \Omega $.

\section{Conclusions}
\label{sec:conclusion}
DM as pNGB, arising out of breaking of a continuous symmetry has
been studied in literature. Similar ideas have also been exploited to
realize composite DM, which indeed appeal to a lot of astrophysical
observations like non-cuspy halo profile etc. (see for example,
\cite{Cline:2013zca,Kopp:2016yji,Ma:2017vzm,Ballesteros:2017xeg,
Balkin:2017yns,Frigerio:2012uc,Antipin:2015xia,Marzocca:2014msa,Khlopov:2008ty}).
Relating this type of DM to a consistent inflationary picture where
the existence of pNGB is an artifact of the breaking of the continuous
symmetry at the end of inflation, is the most interesting feature of
our study.
In this work, we have made use of the pNBGs, which are part of an SQCD framework in realizing early 
Universe inflation, as dark matter candidates. Due to the non-abelian nature of the chiral symmetry 
which was broken spontaneously at the end of inflation, a multiple such pNBGs as dark matter follow 
in the set-up. We have shown that depending on the explicit chiral symmetry breaking term, there 
could be different degree of degeneracy among the masses of these pNBGs. In addition, the presence 
of R-symmetry preserving supersymmetric Standard Model induces in general another candidate of DM, 
called the neutralino (being LSP). Hence we end up having a multi-particle DM scenario, which eventually 
is considered to be dominated by the pNBGs only as far as the contribution to relic density is concerned. 
We then divide our analysis into two parts; one is when the interaction between LSP and pNBGs is completely 
neglected and the other one is with non-zero but small LSP-pNBGs interaction. We find that the case 
with all degenerate pNBGs can not lead to a successful situation consistent with the recent direct detection 
limit in the first case. On the other hand, the case with non-degenerate pNBGs without any 
effective contribution of the LSP toward relic density can be consistent with direct search bound. 
In case of small but non-zero  LSP-pNBG interaction, we have found that this possibility alters our previous 
conclusions significantly. For example, the case of fifteen degenrate pNBG DM now becomes a possibility. 
Therefore our model provides an interesting possibility of pNBG dark matter scenario.

\begin{center}
 \textbf{ACKNOWLEDGEMENT}
\end{center}
SB would like to acknowledge DST-INSPIRE Faculty Award IFA-13 PH-57 at IIT Guwahati. AKS would like to acknowledge MHRD for
a research fellowship.

\section{Appendix I: Dashen Formula} 
Dashen formula for pNGB reads as 
\begin{align}
\langle \chi \rangle^2 (m^a_{G_s})^2&=\langle 0|[\tilde{\mathcal{Q}}_i,[\tilde{\mathcal{Q}}_i,H]]|0\rangle\\
& =\bar{\psi}\Big[\frac{\lambda_a}{2},\Big[\frac{\lambda_a}{2},m_{\textrm{diag}} \Big]_+\Big]_+\psi
\end{align}
where $i=1,2,..4$ and 
\begin{equation}
m_{\textrm{diag}}=
 \begin{pmatrix}
  m_1 & 0 & 0 & 0\\
0 & m_2 & 0 & 0\\
0 & 0 & m_3 & 0\\
0 & 0 & 0 & m_4
 \end{pmatrix}
 \label{eq:mdiag}
\end{equation}
 and $\psi=\{Q_1,Q_2,Q_3,Q_4\}$ is the quark state. $\tilde{\mathcal{Q}_i}=\int d^3 x \psi^\dagger(x)\gamma_5\frac{\lambda_a}{2}\psi(x)$
is the axial charge of the broken $SU(4)$ . $H=\bar{\psi}M\psi$
$\lambda_a$ are the generators of broken $SU(4)_A$ with $a=1,2,....15$. When quark condensates
 like in form of $\langle\bar{Q}Q \rangle=\Lambda^3$, 
we find for a=1,
\begin{align}
&\lambda_1=\begin{pmatrix}
  0 & 1 & 0 & 0\\
  1 & 0 & 0 & 0\\
   0 & 0 & 0 & 0\\
0 & 0 & 0 & 0
 \end{pmatrix}\\
\textrm{Correspondingly,    }
&(m^1_{G_s})^2=(m_1+m_2)\langle\psi\bar{\psi}
\rangle=(m_1+m_2)\frac{\Lambda^3}{\langle\chi\rangle^2}
\end{align}
Similarly using other $\lambda^a$'s for SU(4), we find ($\zeta = \Lambda/\langle\chi\rangle$)
\begin{align}
m^{2,3}_{G_s}&=\zeta \sqrt{(m_1+m_2) \Lambda}\,, \quad
m^{4,5}_{G_s}=\zeta\sqrt{(m_1+m_3) \Lambda}\,, \quad
m^{6,7}_{G_s}=\zeta\sqrt{(m_2+m_3) \Lambda}\\
m^{9,10}_{G_s}&=\zeta\sqrt{(m_1+m_4) \Lambda}\,, \quad
m^{11,12}_{G_s}=\zeta\sqrt{(m_2+m_4) \Lambda}\,, \quad
m^{13,14}_{G_s}=\zeta\sqrt{(m_3+m_4) \Lambda}\\
m^8_{G_s}&=\zeta\sqrt{(m_1+m_2+4m_3)\Lambda^3/3}\,, \quad
m^{15}_{G_s} =\zeta\sqrt{(m_1+m_2+m_3+9m_4) \Lambda^3/6}
\end{align}
%

%%%%%%%%%%%%%%%%%%%%%%%%%%%%%%
\section{Appendix II: pNBG annihilation cross-section to Neutralino}
%%%%%%%%%%%%%%%%%%%%%%%%%%%%%%

 In MSSM, after electroweak symmetry breaking, neutral gauginos and
 Higgsinos mix to yield four physical fields called neutralinos. In the mass basis, the neutralino 
 can be written as a combination of wino, bino and two Higgssions. For example, the lightest one 
 can be written as 
  \begin{align}
  \chi_0=Z_{11}\tilde{B}+Z_{12}\tilde{W}+Z_{13}\tilde{H_u}+Z_{14}\tilde{H_d},
  \end{align}
 where the coefficients $Z_{ij}$ are the elements of the diagonalizing mass matrix and crucially control its interaction to other MSSM and SM particles.
 In our analysis, we have considered that the pNBG's can only annihilate to lightest neutralino by
 assuming the rest to be heavier than the pNBGs. The interaction
 lagrangian of $\chi_0\bar{\chi}_0 h$ vertex is $\mathcal{L}=-i g_2\bar{\chi}_0(C_{L1}P_L+C_{R1}P_R)\chi_{0} h$. Our aim is to 
 calculate the cross section for $G_S G_S\rightarrow \bar{\chi_0} \chi_0 $ annihilation process that has been used in the 
estimation of pNGB DM relic density. It is a two body scattering process and the differential scattering cross section 
in centre of mass frame is given by
\begin{align}
  \frac{d\sigma}{d\Omega}=&\frac{1}{16\pi^2|v_1-v_2|}\frac{p_f}{s^{3/2}}\overline{|M|^2}\label{diffC}\,, \quad p_f=\frac{\sqrt{\{s-(m_3+m_4)^2\}\{s-(m_3-m_4)^2\}}}{2\sqrt{s}}
  \end{align}
 ( $p_f$ is the final state momentum). The Feynman amplitude for the process is
\vspace{2mm}
\begin{align}
-iM=& -i \alpha_2\Big\{C_{L1}v_3\frac{1-\gamma_5}{2}\bar{u_4}+C_{L2}v_3\frac{1+\gamma_5}{2}\bar{u_4}\Big\}\frac{\alpha_1}{(p_2-p_1)^2-m_h^2}
\end{align}
 where $\alpha_2=g_2(C_L P_L+C_R P_R)$ and $\alpha_1=2\lambda^{\prime\prime} v_d$ are two vertex factors in Fig.\ref{fig:fdcon}
as obtained from Eq.(\ref{eq:int}). Using standard procedure, we find
\begin{align}
 \overline{ |M|^2}=\frac{2\alpha_2^2\alpha_1^2}{(s-m_h^2)^2}\Big[(|C_{L}|^2+|C_{R}|^2)\frac{(s-2m_3^2)}{2}-(C_{L}^*C_{R1}+C_{R}^*C_{L})m_3^2\Big]
\end{align}
Finally from Eq.(\ref{diffC}) we obtain the s-wave contribution to the annihilation cross-section as,
\begin{align}
%\sigma =&\frac{1}{4\pi|v_1-v_2|}\frac{p_f}{s^{3/2}}|M|^2\\
\sigma v_{rel}=&\frac{1}{4\pi}\frac{\sqrt{s-4 m_{\chi}^2}}{s^{3/2}}\frac{\alpha_1^2\alpha_2^2}{(s-m_h^2)^2}\Big[(|C_{L}|^2+
|C_{R}|^2)\frac{(s-2m_{\chi}^2)}{2}-(C_{L}^*C_{R}+C_{R}^*C_{L})m_{\chi}^2\Big],
\end{align}
where 
\begin{align}
&C_{L}=-Q_{11}^{\prime\prime*}\sin\alpha-S_{11}^{\prime\prime*}\cos\alpha,\nonumber\\
&C_{R}=-Q_{11}^{\prime\prime}\sin\alpha-S_{11}^{\prime\prime}\cos\alpha,\nonumber\\
&Q_{11}^{\prime\prime}=\frac{1}{2}\Big[Z_{13}(Z_{12}-\tan\theta_WZ_{11})+Z_{13}(Z_{12}-\tan\theta_W Z_{11})\Big],\nonumber\\
&S_{11}^{\prime\prime}=\frac{1}{2}\Big[Z_{14}(Z_{12}-\tan\theta_WZ_{11})+Z_{14}(Z_{12}-\tan\theta_W Z_{11})\Big],\nonumber\\
&m_1=m_2=m_{G_S},\nonumber\\
&m_3=m_4=m_\chi,\nonumber\\
&\sqrt{s}=E_1+E_2=2 E_1=2 m_{G_S}.
\end{align}
For simplification, we have assumed $C_{L}=C_R$ in our analysis.

%\begin{align}
%(m^2_{G_s})^2=(m_1+m_2)\frac{\Lambda^3}{\langle\chi\rangle^2}\\
%(m^3_{G_s})^2=(m_1+m_2)\frac{\Lambda^3}{\langle\chi\rangle^2}\\
%(m^4_{G_s})^2=(m_1+m_3)\frac{\Lambda^3}{\langle\chi\rangle^2}\\
%(m^5_{G_s})^2=(m_1+m_3)\frac{\Lambda^3}{\langle\chi\rangle^2}\\
%(m^6_{G_s})^2=(m_2+m_3)\frac{\Lambda^3}{\langle\chi\rangle^2}\\
%(m^7_{G_s})^2=(m_2+m_3)\frac{\Lambda^3}{\langle\chi\rangle^2}\\
%(m^8_{G_s})^2=\frac{(m_1+m_2+4m_3)}{3}\frac{\Lambda^3}{\langle\chi\rangle^2}\\
%m^9_{G_s})^2=(m_1+m_4)\frac{\Lambda^3}{\langle\chi\rangle^2}\\
%m^{10}_{G_s})^2=(m_1+m_4)\frac{\Lambda^3}{\langle\chi\rangle^2}\\
%(m^{11}_{G_s})^2=(m_2+m_4)\frac{\Lambda^3}{\langle\chi\rangle^2}\\
%(m^{12}_{G_s})^2=(m_2+m_4)\frac{\Lambda^3}{\langle\chi\rangle^2}\\
%(m^{13}_{G_s})^2=(m_3+m_4)\frac{\Lambda^3}{\langle\chi\rangle^2}\\
%(m^{14}_{G_s})^2=(m_3+m_4)\frac{\Lambda^3}{\langle\chi\rangle^2}\\
%(m^{15}_{G_s})^2=\frac{(m_1+m_2+m_3+9m_4)}{6}\frac{\Lambda^3}{\langle\chi\rangle^2}
%\end{align}

%%%%%%%%%%%%%%%%%%%%%%%%%%%%%%%%%%%%
\section{Appendix III: Numerical estimate of Boltzmann equations} 
\label{sec:app.III}
%%%%%%%%%%%%%%%%%%%%%%%%%%%%%%%%%%%%

Relic density allowed parameter space of non-degenerate multipartite DM components of the model (Section 4.1.2 and Section 4.2) has been 
obtained by using approximate analytic solution (Eq.~(\ref{eq:BEQ-nondeg})). Here we will explicitly demonstrate the viability of such analytic 
solution to the exact numerical solution of the coupled Boltzmann equations (BEQ) that defines the freeze-out of such non-degenerate DMs. 
We will illustrate the case of non-degenerate pNGBs, with negligible interactions to LSP (Section 4.1.2). 

Let us define $Y={n_i}/{s}$, where $n_i$ is the number density of $i$'th DM candidate 
 and $s$ is the entropy density of the universe. The BEQ is rewritten as a function of $x={m}/{T}$,
 where $m$ is the mass of DM particle and T is the temperature of the thermal bath. As we have three DM candidates of type A, B and C, 
 we use instead a common variable $x={\mu}/{T}$ where ${1}/{\mu}={1}/{m_A}+{1}/{m_B}+{1}/{m_C}$ and ${1}/{x}={T}/{\mu}=
 {1}/{x_1}+{1}/{x_2}+{1}/{x_3}$. Assuming $m_C>m_B>m_A$, the coupled BEQ then reads
 \begin{align}
  \frac{dY_C}{dx}=&-0.264 M_P \sqrt{g_*}\frac{\mu}{x^2}\Big[\langle\sigma v\rangle_{CC\rightarrow SM}(Y_C^2-Y_C^{\textrm{eq2}})
   +6\langle\sigma v\rangle_{CC\rightarrow B_i B_i}\Big(Y_C^2-\frac{Y_C^{\textrm{eq2}}}{Y_{B_i}^{\textrm{eq2}}}Y_{B_i}^2
  \Big)\nonumber\\+& 8\langle\sigma v\rangle_{CC\rightarrow A_iA_i}\Big(Y_C^2-\frac{Y_C^{\textrm{eq2}}}{Y_{A_i}^{\textrm{eq2}}}Y_{A_i}^2\Big)\Big]\nonumber\\
   \frac{dY_{B_i}}{dx}=&-0.264 M_P \sqrt{g_*}\frac{\mu}{x^2}\Big[\langle\sigma v\rangle_{B_iB_i\rightarrow SM}(Y_{B_i}^2-Y_{B_i}^{\textrm{eq2}})
   -\langle\sigma v\rangle_{CC\rightarrow B_i B_i}\Big(Y_C^2-\frac{Y_C^{\textrm{eq2}}}{Y_{B_i}^{\textrm{eq2}}}Y_{B_i}^2
  \Big)\nonumber\\+& 8\langle\sigma v\rangle_{B_iB_i\rightarrow A_iA_i}\Big(Y_{B_i}^2-\frac{Y_{B_i}^{\textrm{eq2}}}{Y_{A_i}^{\textrm{eq2}}}Y_{A_i}^2
\Big)\Big],\nonumber\\
  \frac{dY_{A_i}}{dx}=&-0.264 M_P \sqrt{g_*}\frac{\mu}{x^2}\Big[\langle\sigma v\rangle_{A_iA_i\rightarrow SM}(Y_{A_i}^2-Y_{A_i}^{\textrm{eq2}})
   -\langle\sigma v\rangle_{CC\rightarrow A_iA_i}\Big(Y_C^2-\frac{Y_C^{\textrm{eq2}}}{Y_{A_i}^{\textrm{eq2}}}Y_{A_i}^2\Big)\nonumber\\
   &-\langle\sigma v\rangle_{B_iB_i\rightarrow A_iA_i}\Big(Y_{B_i}^2-\frac{Y_{B_i}^{\textrm{eq2}}}{Y_{A_i}^{\textrm{eq2}}}Y_{A_i}^2\Big)\Big],
 \end{align}
where the equilibrium distribution has the form 
\begin{align}
Y_i^{\textrm{eq}}(x)=0.145 \frac{g}{g_*}x^{3/2}\Big(\frac{m_i}{\mu}\Big)^{3/2}e^{-x\Big(\frac{m_i}{\mu}\Big)}.
\end{align}

We have already explained that the relic density of A-type DM having mass $\sim\frac{m_h}{2}$ is negligible due to 
resonance enhancement of annihilation cross-section. Therefore, it freezes out much later than B and C type DMs. 
During the freeze out of B and C type DM, we can then safely write $Y_A\simeq Y_A^{\textrm{eq}}$ 
and ignore its contribution to the relic density. The coupled BEQ then effectively turns to
\begin{align}
\frac{dy_C}{dx}=-\frac{1}{x^2}\Big\{ (\langle\sigma v\rangle_{CC\rightarrow SM}+8\langle\sigma v\rangle_{CC\rightarrow A_iA_i})(y_C^2-y_C^{\textrm{eq}^2})
+6 \langle\sigma v\rangle_{CC\rightarrow B_iB_i}(y_C^2-\frac{y_C^{\textrm{eq}^2}}{y_{B_i}^{\textrm{eq}^2}}y_{B_i}^2)\Big\},\nonumber\\
\frac{dy_{B_i}}{dx}=-\frac{1}{x^2}\Big\{ (\langle\sigma v\rangle_{B_iB_i\rightarrow SM}+8\langle\sigma v\rangle_{B_iB_i\rightarrow A_iA_i})(y_C^2-y_C^{\textrm{eq}^2})
-6 \langle\sigma v\rangle_{CC\rightarrow B_iB_i}(y_C^2-\frac{y_C^{\textrm{eq}^2}}{y_{B_i}^{\textrm{eq}^2}}y_{B_i}^2)\Big\},\label{BoltzC2}
\end{align}
where 
\begin{align}
y_i&=0.264 M_P\sqrt{g_*}\mu Y_i,\\
y_i^{\textrm{eq}}&=0.264 M_P \sqrt{g_*}\mu Y_i^{\textrm{eq}}.
\end{align}
\begin{figure}[htb!]
$$
 \includegraphics[height=5.5cm]{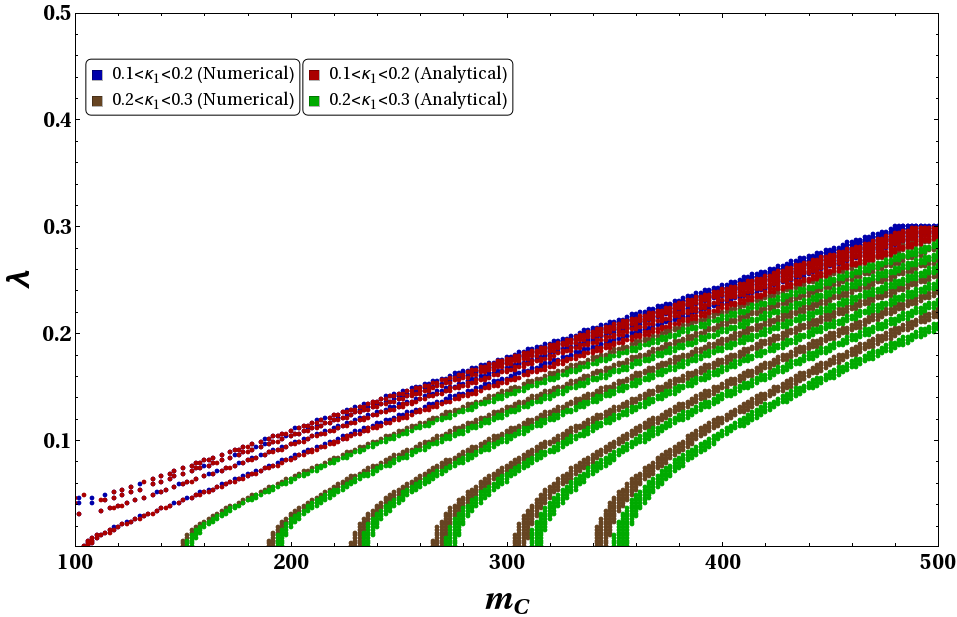}
$$
\caption{Comparison between relic density found using numerical solution to coupled Boltzmann equation (Eq.(\ref{BoltzC2}))
and approximate analytical solutions as in Eq.(\ref{relic:BC}) in $m_C$ (GeV) -$\lambda$ plane for $\kappa:\{0.1-0.3\}$.}
\label{fig:Ana-NuCom}
\end{figure}
Once we obtain the freeze out temperature by solving the set of coupled equations (as in Eq.(\ref{BoltzC2}) numerically, we can 
compute the relic density for each of the DM species by
\begin{align}
\Omega_C h^2&=\frac{854.45\times 10^{-13}}{\sqrt{g_*}}\frac{m_C}{\mu} y_C\Big[\frac{\mu}{m_C}x_{\infty}\Big], \nonumber\\
\Omega_B h^2&=\frac{854.45\times 10^{-13}}{\sqrt{g_*}}\frac{m_B}{\mu} y_B\Big[\frac{\mu}{m_C}x_{\infty}\Big],\\
\Omega_T& \simeq\Omega_C+6\Omega_B.
\end{align}
$y_i\Big[\frac{\mu}{m_i}x_{\infty}\Big]$ indicates the value of $y_i$ evaluated at $\frac{\mu}{m_i}x_{\infty}$, where $x_{\infty}$ denotes 
a very large value of $x$ after decoupling. For numerical analysis we have taken $x=500$ which is a legitimate choice. We scan for $\kappa_1\sim 0.1-0.3$ 
and $m_{C}=100-500$ GeV to find out relic density allowed points. We have shown it in terms of $\lambda-m_C$ in Fig.\ref{fig:Ana-NuCom}. $\kappa:\{0.1-0.2\}$. Analytical solutions (Eq. (\ref{relic:BC})) for relic density is plotted in the same graph for comparative purpose.
 We see for low values of $\kappa:\{0.1-0.2\}$ the numerical solution (in blue) and approximated analytical solution (in red) falls on top of each other with 
 very good agreement. With larger $\kappa:\{0.2-0.3\}$, the separation between numerical (in grey) and approximate analytical solution (in green) increases mildly within $\Delta m_{DM}\sim 10$ GeV.

%%%%%%%%%%%%%%%%%%   References %%%%%%%%%%%%%%%%%%%%%%%%%%%%%%%%%%%%

%\bibliography{z3-2-ref.bib}
%%%%%%%%%%%%%%%%%%%%%%%%%%%%%%%%%%%%%%%%%%%%%%%%%%%%%%%%%%%%%%%%%%%%

\end{document}